\documentclass[%
reprint,
showpacs,preprintnumbers,
nofootinbib,
amsmath,amssymb,
aps,
]{revtex4-1}

\usepackage{graphicx}
\usepackage{dcolumn}
\usepackage{bm}
\usepackage{epstopdf}
\renewcommand{\vec}[1]{\hm{#1}}

\begin{document}

\title{Nonlinear atomic vibrations and structural phase transitions in strained carbon chains}
\author{G.M. Chechin}
  \email{gchechin@gmail.com}
\author{D.A. Sizintsev, O.A. Usoltsev}
\affiliation{South Federal University, Department of physics, Russia}
\date{\today}
\begin{abstract}
We consider longitudinal nonlinear atomic vibrations in uniformly strained
carbon chains with the cumulene structure ($=C=C=)_{n}$. With the aid
of ab initio simulations, based on the density functional theory, we have
revealed the phenomenon of the $\pi$-mode softening in a certain range of
its amplitude for the strain above the critical value $\eta_{c}\approx
11\,{\%}$. Condensation of this soft mode induces the structural
transformation of the carbon chain with doubling of its unit cell. This is
the Peierls phase transition in the strained cumulene, which was previously
revealed in [Nano Lett. {\bf 14}, 4224 (2014)]. The Peierls transition leads
to appearance of the energy gap in the electron spectrum of the strained
carbyne, and this material transforms from the conducting state to
semiconducting or insulating states. The authors of the above paper
emphasize that such phenomenon can be used for construction of various
nanodevices. The $\pi$-mode softening occurs because the old equilibrium
positions (EQPs), around which carbon atoms vibrate at small strains, lose
their stability and these atoms begin to vibrate in the new potential wells
located near old EQPs. We introduced the simple classical model representing
a chain whose particles interact via the Lennard-Jones potential that allows
us to describe quite well properties of the $\pi$-mode softening. Using
this model without any adjustable parameters, we were able to obtain the
value of the critical strain $\eta_{c}$, which coincides with that found
by the ab initio calculations within their accuracy. We also study the
stability of the new EQPs, as well as stability of vibrations in their
vicinity. In previous paper [Physica {\bf D203}, 121(2005)], we proved that
only three symmetry-determined Rosenberg nonlinear normal modes can exist in
monoatomic chains with arbitrary interparticle interactions. They are the
above-discussed $\pi$-mode and two other modes, which we call $\sigma$-mode and $\tau$-mode. These modes correspond to the multiplication of the
unit cell of the vibrational state by two, three or four times compared to
that of the equilibrium state. We study properties of $\pi$-, $\sigma$-
and $\tau$-modes in the chain model with arbitrary pair potential of
interparticle interactions and prove that the critical value $\eta_{c}$
corresponds to the inflection point of this pair potential. The similar
study in the framework of the ab initio approach is difficult by necessity
to have a sufficiently acceptable description of the van der Waals
interactions for atomic vibrations with large amplitudes. The analysis of
the possible condensation of $\sigma$- and $\tau$-modes allows us to
suppose that two new types of carbon chains (besides cumulene and polyyne)
can exist with bond lengths alternation different from that of polyyne.

\end{abstract}
\maketitle

\section{Introduction}

Monoatomic carbon chains can exist in two different modifications. The first
one is polyyne, or carbyne-$\alpha $, representing the chain with
alternating single and triple bonds [chemical structure $(-C\equiv
C-)_{n}$]. The second modification is cumulene, or carbyne-$\beta $,
representing the chain with double bonds [chemical
structure $(=C=C=)_{n}$]. Carbyne chains was claimed to be the
strongest material known at the present time. The synthesis of linear carbon
chains up to 6000~atoms was reported in 2016. Because of many unique
mechanical, physical and chemical properties, which were studied or only
predicted, carbyne is considered as perspective material for various
nanodevices, for hydrogen storage, etc. (see~\cite{NanoLett2014, Carb2014, Nano2013, NanoLett2011, PRB2010} and papers cited
therein).

Chemical synthesis of pure carbyne chains and their experimental study are
very difficult and, therefore, theoretical ab initio investigations, in
particular those based on the density functional theory (DFT), play a rather
important role for prediction of its properties and for treating different
physical phenomena, which are possible in this material. Many interesting
results on strained carbyne chains were obtained with DFT computer
simulations. In the paper~\cite{PRB2010}, DFT {\it ab initio} methods allow to reveal that
distribution of bond length and magnetic moments at atomic sites exhibit
even-odd disparity depending on the number of carbon atoms in the chain and
on the type of saturation of these atoms at both ends. It was also found
that a local perturbation created by a small displacement of the single
carbon atom at the center of a long chain induces oscillations of atomic
forces and charge density, which are carried to long distances over the
chain.

In the paper~\cite{NanoLett2014}, the structural transformation of cumulene under a certain
strain was revealed. This is the Peierls phase transition, which leads to
the radical change of carbyne electron spectrum. As a result of this
transition, an energy gap in the electron spectrum appears and the
conductive cumulene transforms into polyyne which is semiconductor or
insulator. This phenomenon opens perspectives to control electrical behavior
of carbyne by mechanical strain~\cite{NanoLett2014}.

During the study of different types of nonlinear atomic vibrations in
strained cumulene chains, under periodic boundary conditions, we revealed an
unexpected phenomenon of softening~of the longitudinal $\pi $-mode
vibrations above a certain critical value of the strain. Some results of
this study were published in the brief paper~\cite{Lett2016}. They can be summarized as
follows\footnote{Unfortunately, we were not aware of the paper~\cite{NanoLett2014} when
prepared our own paper~\cite{Lett2016}.}.

For strains lower than $\eta =11\,{\%}$ cumulene demonstrates monotonic
{\it hard} type of nonlinearity (the frequency increases with increasing the $\pi$-mode amplitude $a$). However, for $\eta >11\,{\%}$ there is a certain range of amplitudes $a$ in which {\it soft} nonlinearity occurs, namely, the frequency of the
$\pi$-mode abruptly {\it decreases} and then again begins to increase.

The phenomenon of vibrational modes softening is well known in the theory of
structural phase transitions~\cite{PRL1959} where by condensation (``freezing'') of
such modes one tries to explain the nature of the displacement-type phase
transitions. This is the so-called concept of {\it soft modes}. It is essential that in the
majority of the papers on this subject, soft modes are treated in purely
phenomenological manner with some vague arguments about changing of
electron-phonon interactions in crystal under change of such external
parameters as temperature and pressure. Unlike these works, in our study a
soft vibrational mode in cumulene appears as a direct result of the ab
initio simulation {\it without} any additional assumptions.

In~\cite{Lett2016}, the phenomenon of the $\pi$-mode softening has been explained by
the fact that above the critical value of the strain the old atomic
equilibrium positions (EQPs) become unstable and two new EQPs appear near
each of them. Namely, vibrations in the vicinity of these new EQPs
correspond to the softening of the $\pi$-mode. In turn, condensation of the
$\pi$-mode leads to a new atomic equilibrium configuration that corresponds
to the Peierls phase transition. After this transition, the unit cell turns
out to be twice as large than that of cumulene, and the carbon chain
transforms into another carbyne form, polyyne, with bond lengths
alternation.

The main ab initio results obtained in our work and in the paper~\cite{NanoLett2014} are
sufficiently close to each other. Some discrepancy can be explained by
different approximations used in the framework of DFT approach (different
exchange-correlation functionals, different sets of basis functions for
solving Kohn-Sham equations, different realization of the numerical
methods in the packages ABINIT and VASP, etc.). Nevertheless, our results
and those from~\cite{NanoLett2014} are identical qualitatively (detailed comparison will be
presented elsewhere).

In the present paper, we discuss in detail the condensation of the $\pi
$-mode as well as the condensation of two other symmetry-determined
nonlinear normal modes, which are possible in cumulene chains. With the aid
of our approach combined with some group-theoretical methods \cite{PhysD2005}, we
predict the possibility of existence of two new types of carbon chains,
besides cumulene and polyyne. They both possess alternation of bond lengths,
but with {\it different alternating schemes} compared to that of the polyyne.

This paper is organized as follows. In Sec.~\ref{DFTmodel}, we consider the $\pi $-mode
atomic vibrations in the strained cumulene in the framework of the DFT model
and discuss the properties of these vibrations near new EQPs. The simple
model of the monoatomic chain whose particle interact via the Lennard-Jones
potential (L-J~chain) is introduced in Sec.~\ref{LJ_sec} and the appearance of the
new EQPs is explained. In Sec.~\ref{sec_Stability}, we discuss the stability of these
equilibrium positions. In Sec.~\ref{Ros_sec}, the notion of the symmetry-determined
Rosenberg NNMs in monoatomic chains is considered. In Sec.~\ref{LJ_NNM_sec}, we discuss the
properties of these modes in the framework of the L-J model and results of
their condensation in the cumulene chains. Sec.~\ref{Conclus_sec} contains some conclusion
remarks on nonlinear dynamics of the strained carbon chains.

\section{{\large $\pi$}-mode vibrations within DFT model \label{DFTmodel}}

We investigate {\it longitudinal} atomic vibrations of {\it uniformly strained} carbon chains in the $\pi$-mode
dynamical regime. This strain is modeled by an artificial increase of the
unit cell size ($R)$ with respect to that of the chain without strain
($R_{0})$. Thus, speaking about the strain of the chain by $\eta $ per cent,
we mean that $R  =  R_{0}$ ($1+ \eta)$.

The set of atomic displacements $\vec{X} \left( t \right)=[x_{1}\left( t
\right),\, x_{2}\left( t \right)\mathellipsis x_{N}(t)]$ of the
$N${-}particle carbon chain for such vibrational regime at a fixed time
$t=t_{0}$ can be written as follows:

\begin{equation*}\label{eq1}
    \vec{X} (t_{0})=[ a(t_{0}),-a( t_{0})\left| a\left( t_{0} \right),-a( t_{0})
\right|\mathellipsis
\end{equation*}
\begin{equation}
\mathellipsis
\vert a(t_{0}),-a(t_{0})].
\end{equation}
In this pattern, all $x_{i}\left( t \right)$ with odd numbers are equal to
$a\left( t_{0} \right)$, while those with even numbers are equal to
$-a\left( t_{0} \right)$. Thus, all neighboring atoms vibrate out of phase
with equal amplitudes. The unit cell for describing $\pi $-mode vibrations
of the chain with periodic boundary conditions is twice larger than that of
the equilibrium state. Since two carbon atoms in this unit cell possess, at
any time $t$, displacements $x(t)$ and $-x(t)$, one can discuss the time evolution of
only one of them choosing the origin at its equilibrium position. To excite
the $\pi $-mode vibrations in the chain we assume $x(0) =  a$, $\dot{x}(0) = 0$.

The $\pi $-mode {\it nonlinear }oscillations in the strained cumulene we study with the aid of
DFT theory \cite{RMP1999} using for this propose the software package ABINIT \cite{PhysCommon2009,abi}.

The Born-Oppenheimer approximation was used to separate fast motion of
electrons and slow motion of nuclei. At each time step for fixed positions
of nuclei, self-consistent electron density distribution is calculated by
solving Kohn-Sham quantum-mechanical equations. Then forces acting on the
nuclei are computed, and their new configuration is found by using one step
of solution of classical dynamical equations. For this new configuration,
the procedure of self-consistency for the electron subsystem is repeated.

All calculations are performed in the framework of the local density
approximation (LDA). Pseudo-potentials by Troullier-Martins was used to
describe the field of the carbon atoms inner shells in the process of the
Kohn-Sham equations solving with the aid of the plane waves basis (energy
cutoff is equal to 1360~eV). The convergence for energy is chosen as
$10^{-8}$~eV between two steps.

Hereafter, considering ab initio simulations, we speak about the study of
nonlinear oscillations in the framework of the {\it DFT model} emphasizing that our
calculations provide only an approximation to the real physical picture.

\begin{figure}[h!] \centering
\includegraphics[width=1\linewidth]{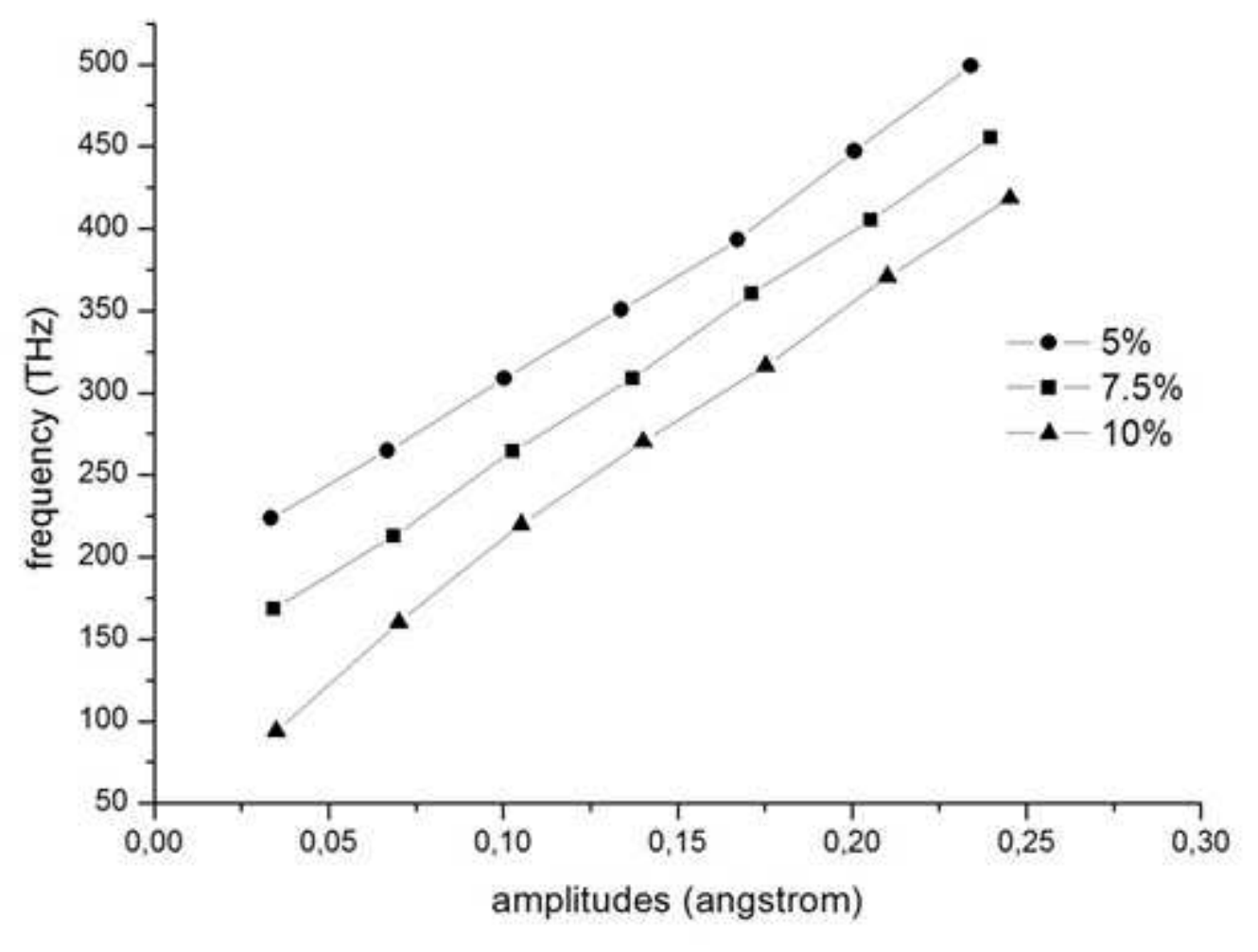} \caption{Dependence of the frequency ($\omega$) of the $\pi$-mode on its amplitude ($a$) for small strains of the carbyne chain.} \label{fig1}
\end{figure}

An important feature of nonlinear vibrations is the dependence $\omega
(a)$ of the frequency $(\omega )$ on the amplitude $(a)$. To find this
dependence, we carried out a series of {\it ab initio }calculations. For each computational
run, with fixed values of the strain and the amplitude (parameter) $a$ of the
$\pi$-mode, the frequency $\omega(a)$ was calculated. In Fig.~\ref{fig1}, we present the function $\omega (a)$ for chains
with $5\,{\%}$, $7,\!5\,{\%}$ and $10\,{\%}$ strain. This figure demonstrates that {\it hard type of nonlinearity}
appears at relatively small strains (increase of amplitude results in
increase of frequency).

\begin{figure}[h!] \centering
\includegraphics[width=1\linewidth]{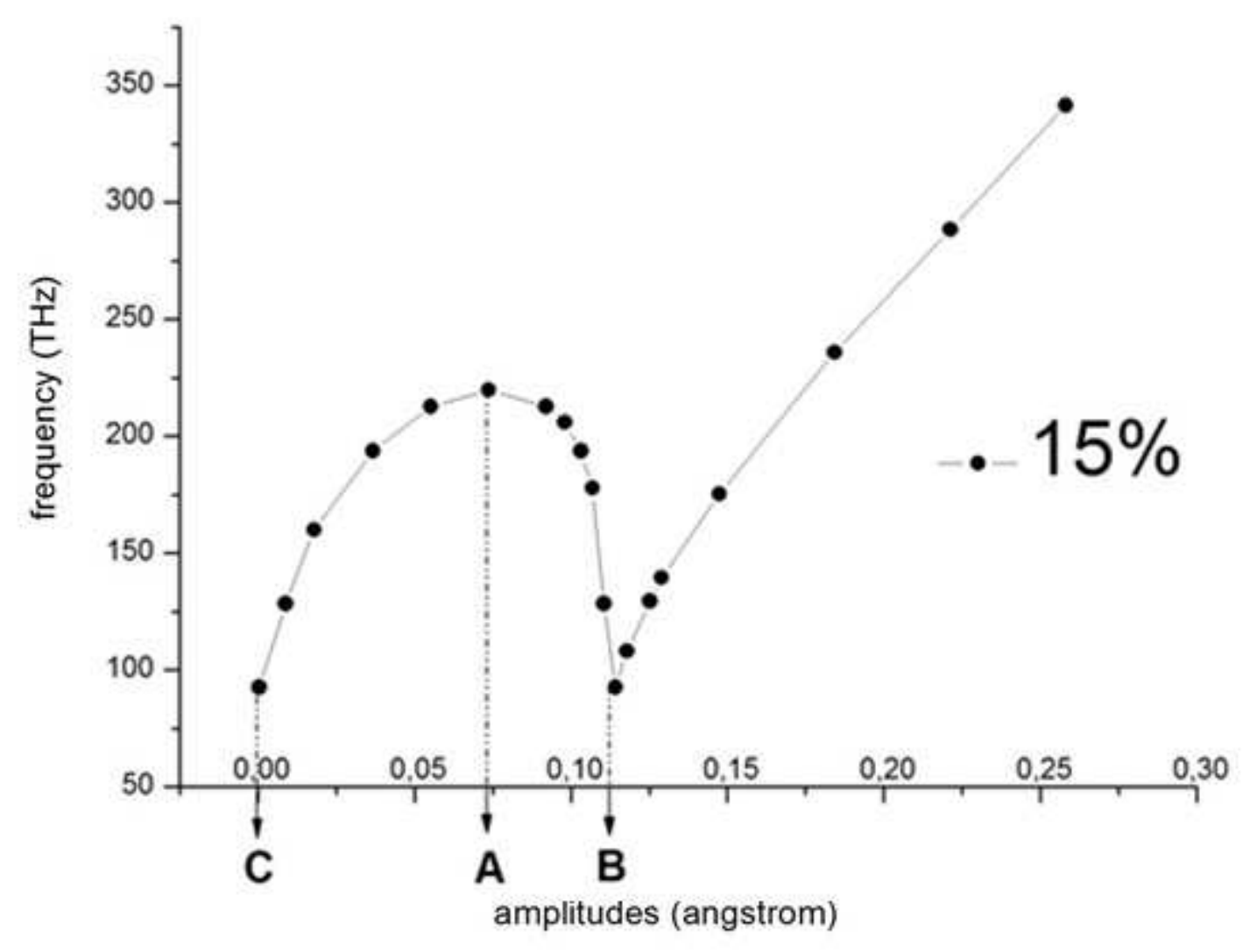} \caption{Dependence of the frequency ($\omega$) of the $\pi$-mode on its amplitude ($a$) for $15\,\%$ strain  of the carbyne chain.} \label{fig2}
\end{figure}

However, we have revealed unexpected behavior of the function $\omega (a)$
for the strain above the critical value $\eta_{c}=11\,{\%}$. For example,
one can see in Fig.~\ref{fig2} that this function for $\eta =15\,{\%}$ turns out to be
nonmonotonic and {\it softening} of the $\pi$-mode, at a certain interval of the parameter
$(a)$, takes place.

\begin{figure}[h!] \centering\includegraphics[width=1\linewidth]{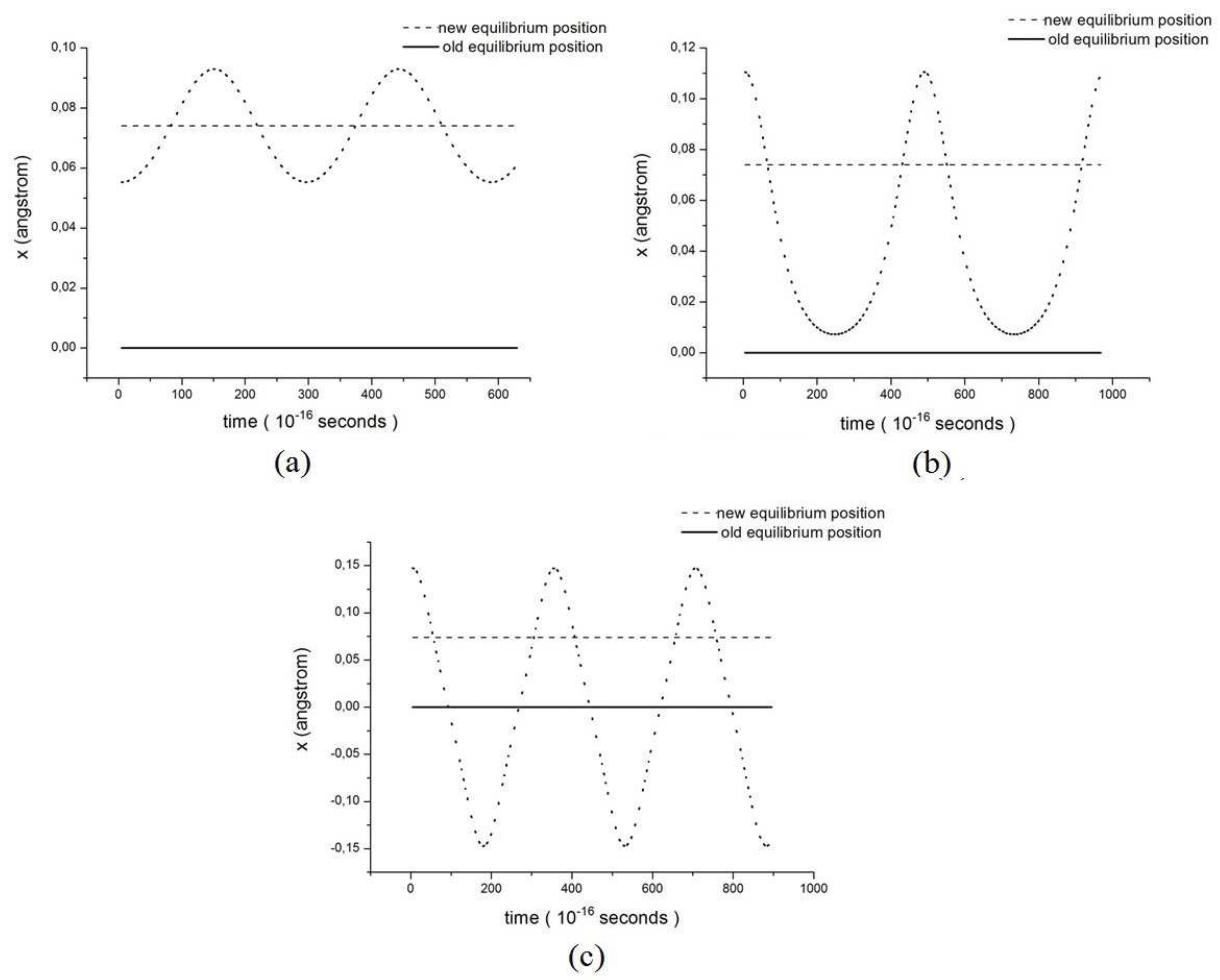} \caption{Oscillations of the carbon atom for different $\pi$-mode amplitudes: (a) oscillations in the small potential well near the new equilibrium position; (b) oscillations before the escape from the small potential well; (c) oscillations in the large potential well with respect to the old equilibrium position.} \label{fig3}
\end{figure}

It is essential that for the $\pi$-mode parameter $a$ belonging to the
interval [{\bf A}, {\bf B}] (Fig~\ref{fig2}) the carbon atom oscillates about a {\it new equilibrium position }different
from the old one at $x=0$. This effect is illustrated in Fig.~\ref{fig3} for the chain
strained by $15\,{\%}$ (detail discussion of these oscillations is given below
in Sec.~\ref{sec_Stability}).

To explain the above behavior of the function $\omega (a)$ one can study the
potential energy\footnote{We excite the $\pi$-mode vibrations by assigning to
all carbon atoms the same displacements $(a)$ and zero velocities at
initial time. Thus, the total energy $E(a)$ of the carbyne chain at this instant
is equal to its potential energy $U(a)$.} of the $\pi$-mode $U(a)$ as a function of
its parameter $a$ for different strains of the carbon chain. For this purpose,
we fix the configuration of carbon nuclei choosing a concrete value of the
parameter $a$, and then find the potential energy $U(a)$ of this configuration with
the aid of the software package ABINIT.

\begin{figure}[h!] \centering
\includegraphics[width=1\linewidth]{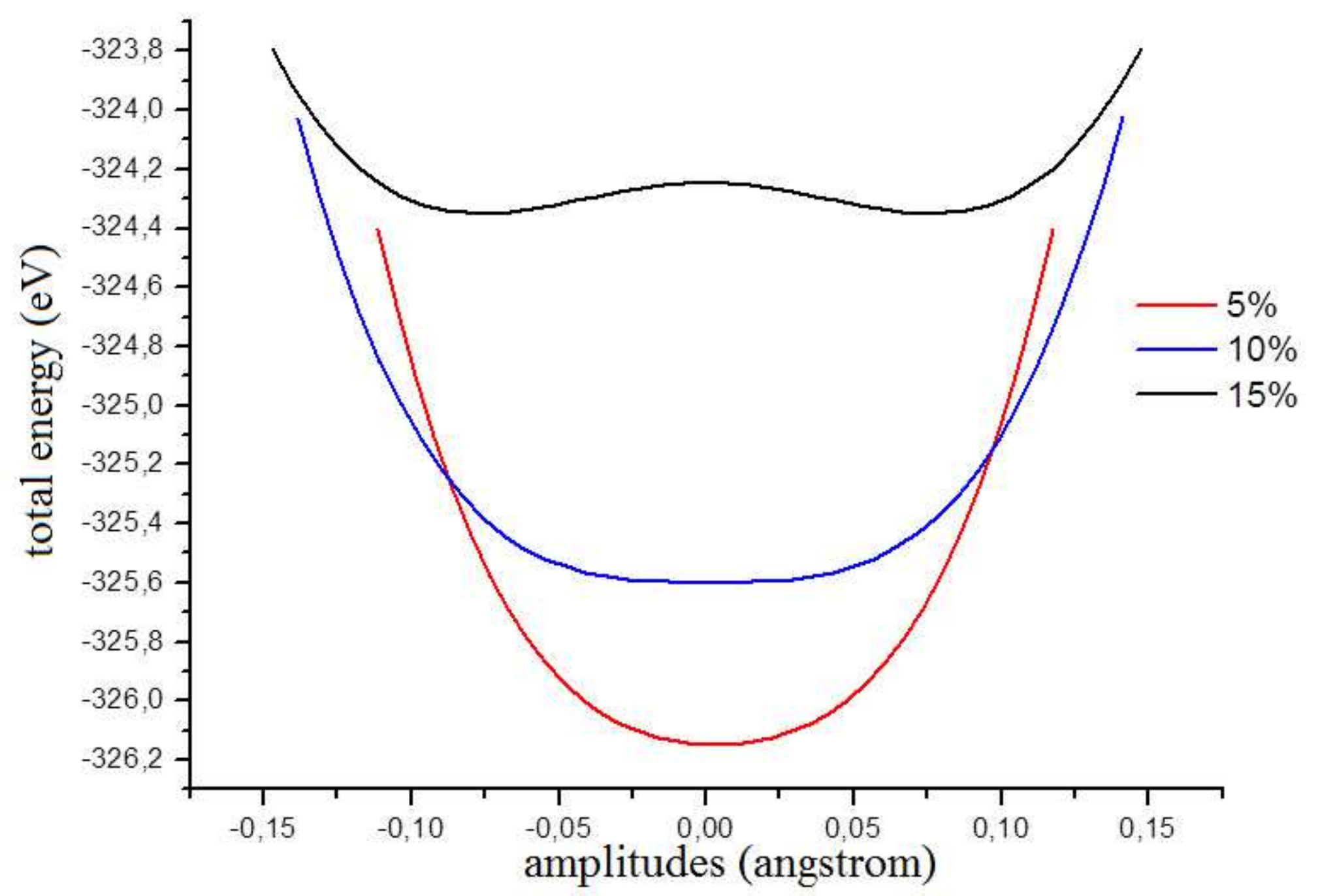} \caption{Profiles of the potential energy of cumulene chains for different strains (color online).} \label{fig4}
\end{figure}

The energy profiles $U(a)$ for different strains of carbon chains are shown in
Fig.~\ref{fig4}. From this figure, one can see that with increase of the strain, the
energy profiles $U(a)$ become flatter near the origin ($a = 0$) and their
curvature changes sign after passing through zero. As a result, the {\it old}
equilibrium position ($a = 0$) becomes unstable, and two {\it new} minima appear at
equal distances on both sides of the origin. This specific behavior of
potential energy of strained carbon chains helps us to explain the properties
of the amplitude-frequency dependence $\omega (a)$ shown in Fig.~\ref{fig2}. Let us
consider this question in more detail.

\begin{figure}[h!] \centering
\includegraphics[width=1\linewidth]{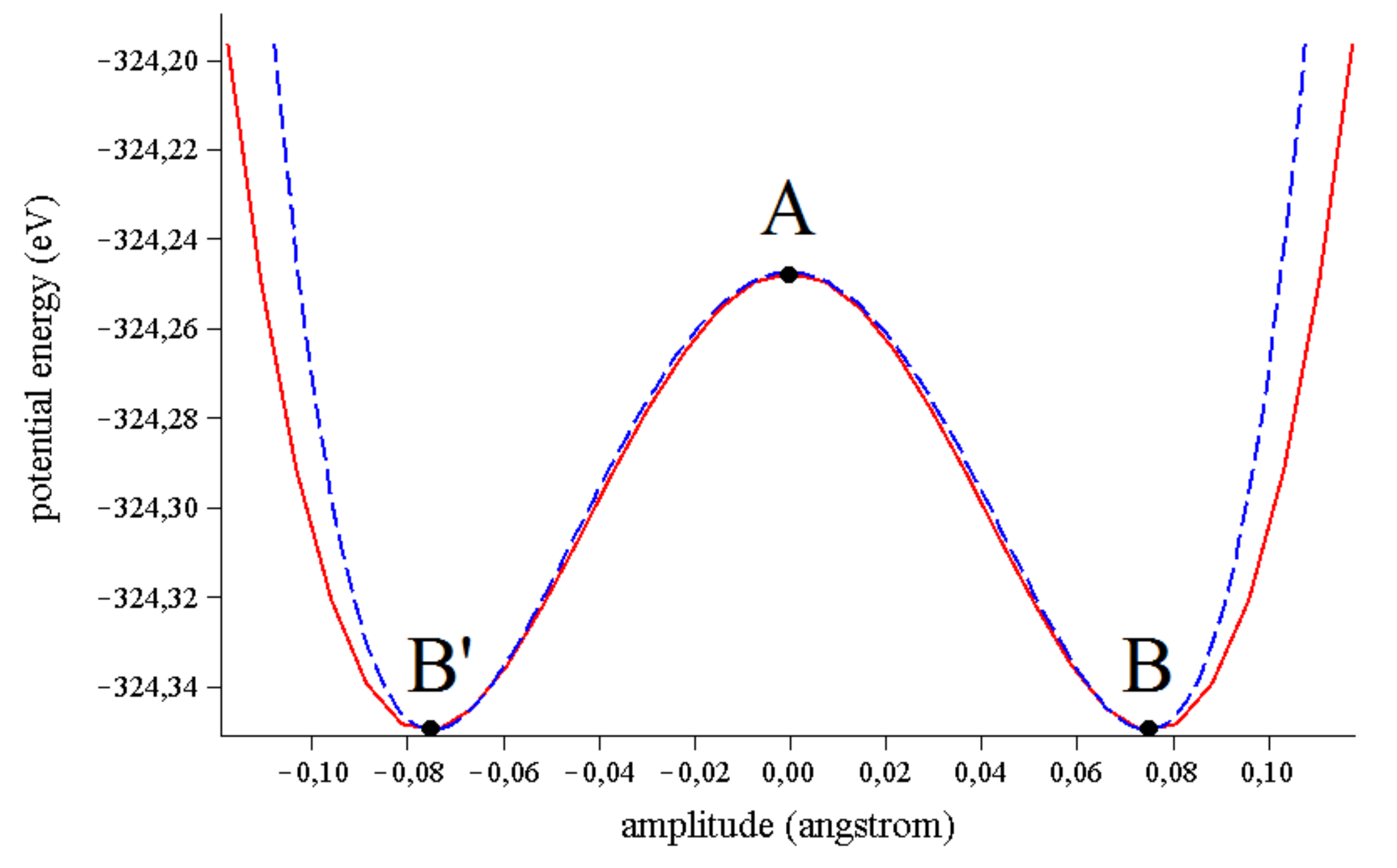} \caption{The $\pi$-mode energy profile for the strain $\eta =15\,{\%}$. The points A and B ($B'$) correspond to the old and new equilibrium positions of carbon atom, respectively. The solid curve shows the ab initio results, while the appropriate fitting in the framework of the Lennard-Jones model is shown by the dashed curve.} \label{fig5}
\end{figure}

In Fig.~\ref{fig5}, one can see the $\pi$-mode potential energy profile for the
strain $\eta =15\,{\%}$, as well as some fitting of this profile in the
framework of the Lennard-Jones model, which is discussed in the next
section. The points A and B ($B'$) in this figure correspond to the old and
new equilibrium positions, respectively.

\begin{figure}[h!] \centering
\includegraphics[width=1\linewidth]{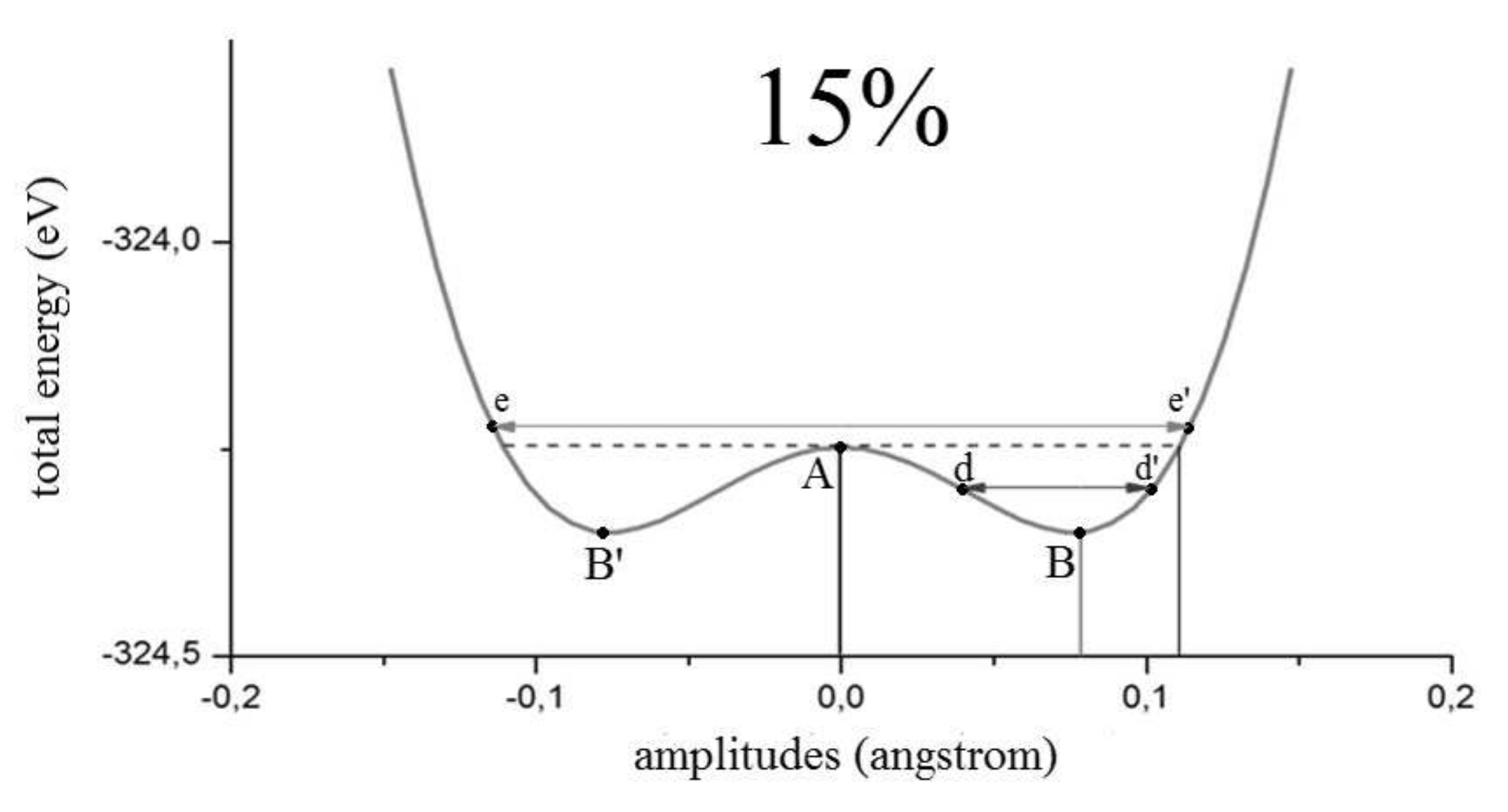} \caption{The potential energy profile of the carbyne chain under $15\,\%$ strain and the $\pi$-mode oscillations of carbon atom with respect to the new (point B) and old (point A) equilibrium positions.} \label{fig6}
\end{figure}

Next, we study the atomic vibrations in the potential wells depicted in
Figs.~\ref{fig5} and~\ref{fig6}.

Let us consider the $\pi $-mode energy $E_{0}=u(a_{0})$ that corresponds
to the new equilibrium position $a=a_{0}$, i.e. to the bottom B of the right
potential well in Fig.~\ref{fig5}. The frequency $\omega (a)$ of the atomic
oscillations near the point $B$ (for example, between returning points $d$ and
$d'$ in Fig.~\ref{fig6}) {\it decreases} with increasing of the energy $E$ from $E_{0}$ up to the value
E$_{c}$ which is determined by the top of the potential hill at $a=0$. Thus
vibrations in the ``small'' potential well demonstrate the {\it soft} type of
nonlinearity.

There is a certain gap in the function $\omega (a)$ for further increase of
$E$, because the carbon atom begin to oscillate in the ``large'' potential
well (for example, between points $e$ and $e'$ in Fig.~\ref{fig6}) around the {\it old} equilibrium
position at $a=0$. The increasing of the energy E of the $\pi$-mode in this
well leads to increasing of the frequency $\omega(a)$. Such behavior of the
frequency corresponds to the {\it hard} type of nonlinearity.

Thus, the phenomenon of the $\pi$-mode softening can be explained by the
atomic vibrations near the new equilibrium positions in small potential
wells of the strained chain. This phenomenon will be considered in more
detail in the framework of the Lennard-Jones model in the next section.

Comparing the above described behavior of the frequency $\omega(a)$ with
that depicted in Fig.~\ref{fig2}, one has to take into account that the amplitude of
the atomic oscillations near the old equilibrium positions at $a=0$ coincides
with the $\pi$-mode parameter $a$, while the vibrational amplitude in the
small potential well near new equilibrium positions is equal to
($a-a_{0})$. This is the reason why we prefer to call $a$ by the term ``$\pi
$-mode parameter'' rather than ``$\pi $-mode amplitude''.

\begin{figure}[h!] \centering
\includegraphics[width=1\linewidth]{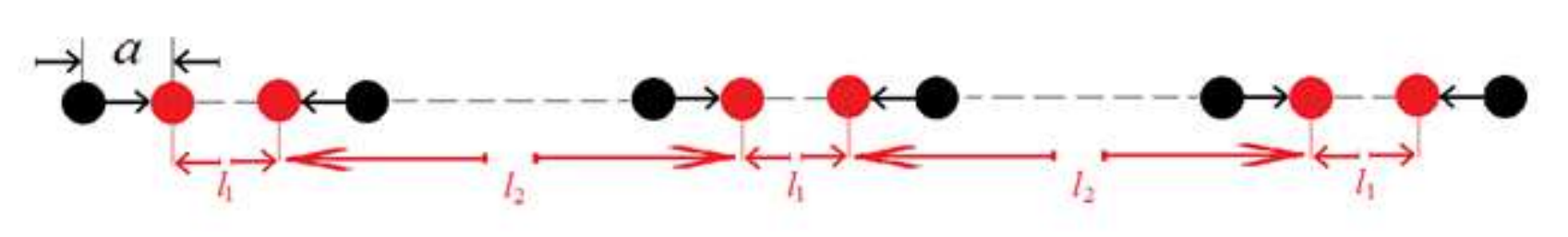} \caption{Bond lengths alternation in strained carbyne for the pattern (1) with $a(t_0 )=a$. Here, for $a>0$ values $l_1$ and $l_2$ are short and long bonds, respectively, and vice versa for $a<0$ (color online). } \label{fig7}
\end{figure}

The condensation of the $\pi $-mode leads to appearance of two domains of
the structural phase transition of the displacement type. If all carbon
atoms of the chain are located in the right potential wells, we have one
domain, while another domain corresponds to the localization of the atoms in
the left wells. In both domains, there is a bond length alternation (BLA)
which can be seen in Fig.~\ref{fig7}, where
$l_{1}=l_{0}-a_{0}$ and $l_{2}=l_{0}+a_{0}$ are short and long
interparticle distances (bond length) for $a_{0}> 0$ and vice versa for
$a_{0}< 0$. From Fig.~\ref{fig7}, we find the following BLA for the $\pi$-mode
condensation:

BLA$_{\pi }=[l_{1}, \ l_{2}| l_{1}, \ l_{2}| ... | l_{1}, \ l_{2}]$.

This BLA corresponds to the polyyne structure of the carbon chain.

\begin{figure}[h!] \centering
\includegraphics[width=1\linewidth]{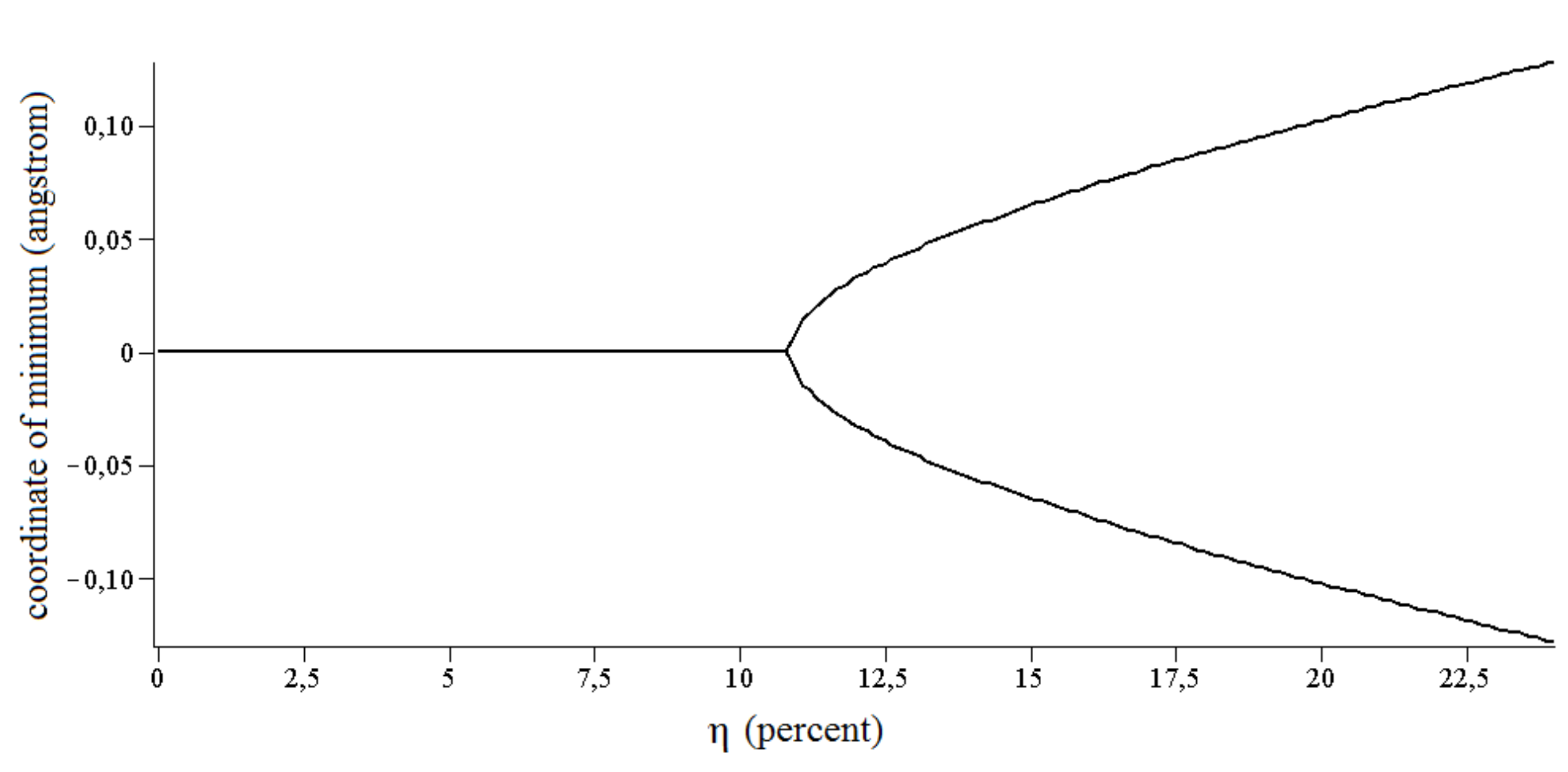} \caption{Dependence of the $\pi$-mode parameter $a_0$ on the strain $\eta$, which corresponds to the minimum of the potential energy (point B in Fig.~\ref{fig5}.). } \label{fig8}
\end{figure}

The transformation from cumulenne to polyyne structure can be observed as a
result of the appropriate increasing of the magnitude $\eta $ of the strain
above the critical value $\eta_{c}$ when the old equilibrium positions lose
stability and two new ones appear near each of them. In Fig.~\ref{fig8}, we present
the dependence $a_{0}(\eta)$ of the position $a_{0}$ of the new equilibrium
state on the strain $\eta $. The bifurcation takes place approximately at
$\eta_{c}=11\,{\%}$.

\newpage

\section{The Lennard-Jones model  \label{LJ_sec}}

Now we consider the Lennard-Jones model, which is associated with the
molecular dynamics approach. The main idea of the molecular dynamics can be
formulated as follows. Molecules (atoms) are replaced by mass points whose
interactions are described with the aid of some phenomenological potentials.
For the obtained dynamical system, equations of classical mechanics are
solved. In the framework of quantum mechanics, such approach cannot be
considered as sufficiently adequate, because it is difficult or impossible
to find potentials which are good enough to take into account the influence
of atomic electron shells on dynamical properties of the original physical
system. That is why one has to use very complicated many-particle potentials
which possess different forms for different geometry of the interacting
atoms and contain phenomenological constants defined by huge tables (see,
for example, \cite{PRB1990}).

However, in some cases, reasonable results can be obtained even with the aid
of simple pair potentials, such as those by Morse, Lennard-Jones, etc. For
example, one can refer to the paper \cite{Let2014} devoted to study discrete
breathers in 2D and 3D Morse crystals. Below, we try to explain the
above-discussed results, obtained by {\it ab initio} calculations for the $\pi $-mode
dynamics in carbyne, using the model of a chain whose particles interact via
the Lennard-Jones potential (L-J chain).

Let us remind some well-known properties of this potential which can be
written in the form
\begin{equation}
\label{eq2}
\varphi(r)=\frac{A}{r^{12}}-\frac{B}{r^{6}}.
\end{equation}
Here the first term describes repulsion between two particles that are at
distance r from each other, while the second term describes their
attraction. The space dependence of this attraction can be explained in the
framework of quantum mechanics as a result of the induced dipole-dipole (van
der Waals) interaction, while the 12-th degree of the distance r in
repulsive term of Eq.(\ref{eq2}) is introduced only for computational convenience.
An important feature of the Lennard-Jones potential is that its both
constants, A and B, can be chosen equal to unity ($A=B=1$) without loss of
generality, if we make an appropriate scaling of space and time variables in
dynamical equations constructed with the aid of this potential.

The dependence of the $\pi$-mode energy on the parameter $a$ was shown in Fig.~\ref{fig5},
where the solid line corresponds to the results of {\it ab initio} calculations, while the dashed line is used for the L-J model. We use the simplest
fitting for the energy profile corresponding to the L-J model. It was
obtained by such choice of two Lennard-Jones parameters, A and B from Eq.(\ref{eq2}),
that the energy at the points A and B (see Fig.~\ref{fig5}) coincides with the energy
of the profile found by {\it ab initio} calculations. Note that Fig.5 corresponds to the
strain $\eta =15\,{\%}$ and for this case $a_{0}$ is equal to $0,\!075{\AA}$.

In Fig.5, one can see considerable discrepancy between the energy profile
u($a)$, which was found by ab initio calculations and its Lennard-Jones fitting
in the case of large values of the $\pi $--mode parameter $a$. The fitting
based on the Morse potential is slightly better, but the above discrepancy
also turns out to be considerable. Certainly, one can set a goal to find
such pair potential that provides more satisfactory fitting, but this goal
is not too important. The matter is that the function u($a)$, in the framework
of the ABINIT model, can be found only with low accuracy for large values of
the $\pi $--mode parameter $a.$ Indeed, such values correspond to large
amplitudes of atomic vibrations and one has to apply a rather accurate
expression for the van der Waals interactions. On the other hand, these
interactions are essentially nonlocal and cannot be described correctly in
the framework of the conventional DFT theory, at least with the aid of the
approximations used in such software packages as ABINIT, VASP, etc.. We
discuss this problem in the last section of the present paper.

We verified that dynamical properties of the $\pi $-mode obtained by the {\it ab initio}
study (see Fig.~\ref{fig2}) and those obtained by molecular dynamics for the L-J chain
are in qualitative agreement. The above facts seem to be important since
computer experiments, based on the density functional theory, require very
long time in contrast to those based on the methods of molecular dynamics.
Therefore, one can use the L-J chain modeling to predict some dynamical
properties of the carbyne, which then may be verified by {\it ab initio} calculations.

It is very interesting to clear up the nature of evolution of the carbyne
properties with increasing of the strain. Why old equilibrium positions lose
their stability and new stable equilibrium positions appear? This problem
can be understood with the aid of the following simple L-J model.

\begin{figure}[h!] \centering
\includegraphics[width=1\linewidth]{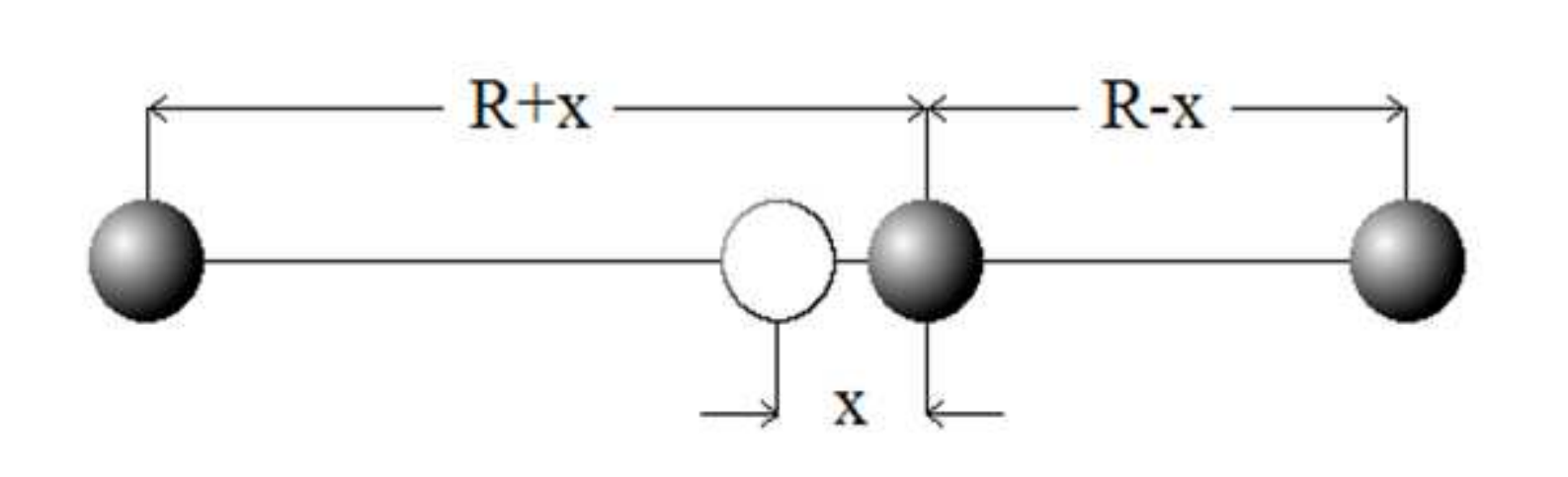} \caption{Simple model of three atoms interacting via the Lennard-Jones potential. Atoms 1, 3 are fixed, while atom 2 is displaced by $x$ from its old equilibrium position at origin. } \label{fig10}
\end{figure}

Let us consider a system of three carbon atoms located at a distance R from
each other symmetrically about the origin $x=0$ (see Fig.~\ref{fig10}). Positions of
the atoms 1 and 3 are fixed, while the ``inner'' atom 2 can be displaced by a
certain value $x$ from its old equilibrium position at origin. We assume that
atoms interact via the Lennard-Jones potential $\varphi (x)$ from Eq.(\ref{eq2}) with
$A=B=1$ (hereafter, it will be called the standard L-J potential).

\begin{figure}[h!] \centering
\includegraphics[width=1\linewidth]{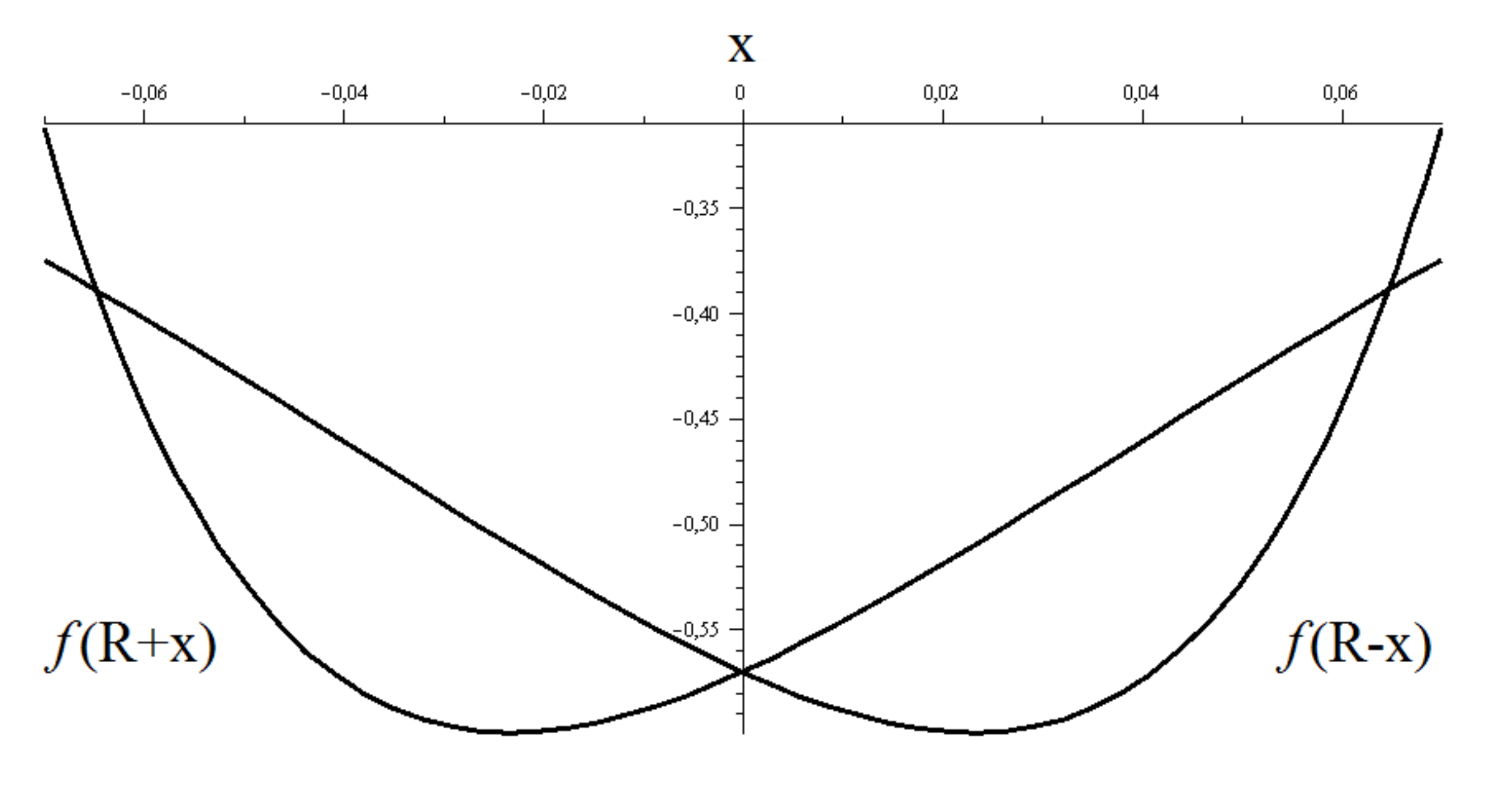} \caption{Appearance of the new equilibrium positions in the simple model of three particles interacting via the Lennard-Jones potential. } \label{fig11}
\end{figure}

If the coordinate x corresponds to the equilibrium state of the atom {\#}2,
then the forces $f\left( r \right)$ acting on it from the left and right
neighbors must be equal:
\begin{equation}
\label{eq3}
f( R+x )= f( R-x ),
\end{equation}
where
\begin{equation}
\label{eq4}
f(r)=-\frac{\mathrm{d}\varphi
}{\mathrm{d}r}=\frac{1}{r^{12}}-\frac{1}{r^{7}}.
\end{equation}
Eq.(\ref{eq3}) represents a nonlinear equation with respect to x that may have
{\it several} real roots. We depict this situation in Fig.~\ref{fig11} where intersection points of the
functions $f(R+x)$ and $f(R-x)$ at $x=0$ and at $x= 0,\!065$ take
place. Here $R=1,\!291$ corresponds to the $15\,{\%}$ strain of the carbyne chain.
The root $x=0$ is associated with old equilibrium position that becomes
unstable, while the roots $x= 0,\!065$ are associated with the new
ones. Thus, the cause of the appearance of the new equilibrium positions
becomes obvious.

It is also easy to verify that the curvature of the potential energy of the
considered system is negative for $R=1,\!20$ (stable equilibrium), while it
becomes positive for $R=1,\!25$ (unstable equilibrium). Note that one can
reveal the similar behavior of the potential energy for L-J chains with
arbitrary number of atoms.

\begin{figure}[h!] \centering
\includegraphics[width=1\linewidth]{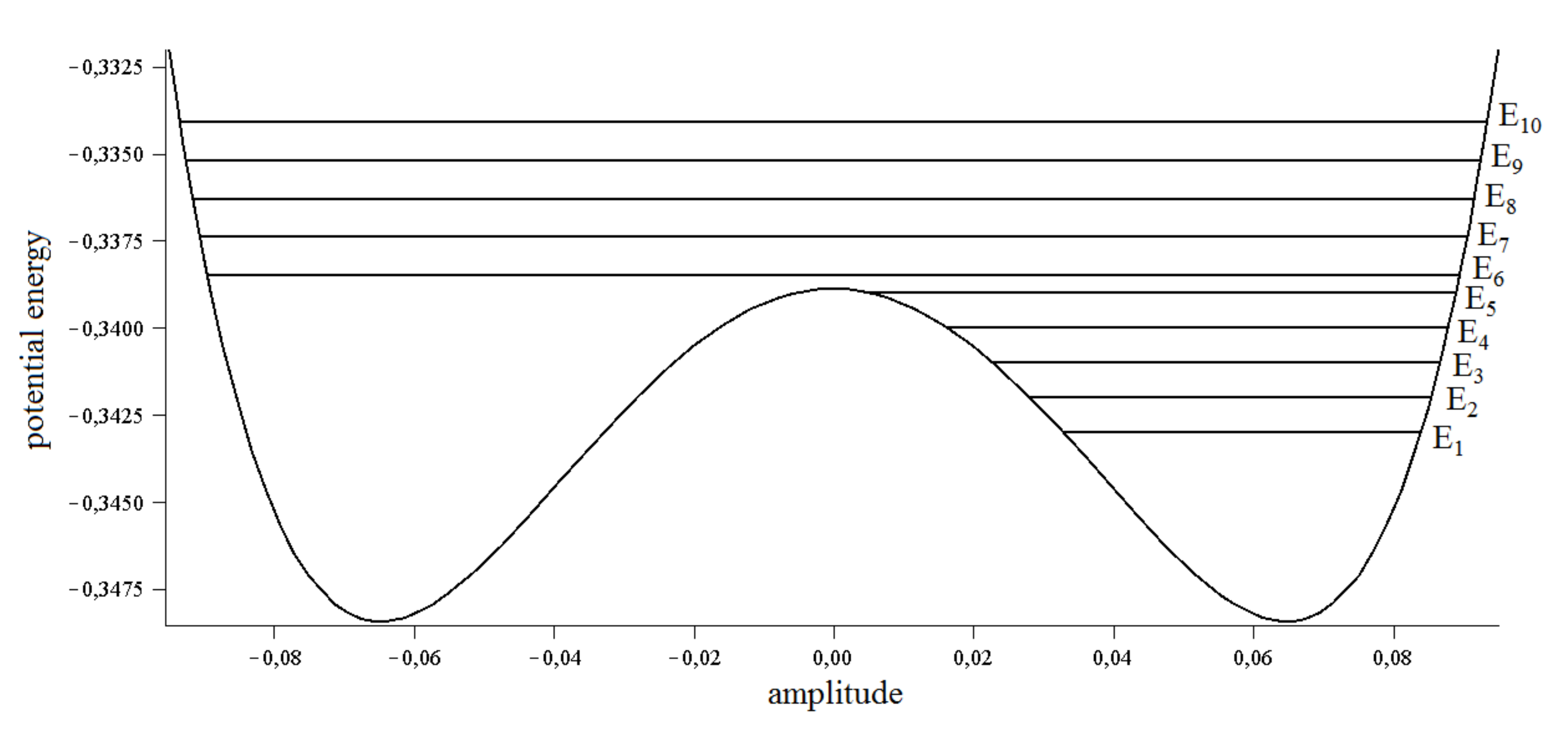} \caption{Some energy levels in the potential well for the Lennard-Jones model, which correspond to the values shown in Table~\ref{t1}. } \label{fig12}
\end{figure}

It is interesting to study nonlinear atomic vibrations in the potential well
corresponding to the standard L-J model (this well is similar to that
depicted in Fig.6 for the ABINIT model). In Fig.~\ref{fig12}, we show some arbitrary
chosen energy levels in the small right well near the new equilibrium
position, as well as levels above the potential hill which correspond to
vibrations in the large potential well around the old equilibrium position
($a=0$). The frequencies corresponding to these levels are presented in Table
1. From this table one can see the soft nonlinearity of vibrations near the
new EQPs in the small right potential well (the frequency decreases with
increasing of the energy) and the hard nonlinearity above potential hill in
the large well (the frequency increases with increasing of the energy).
Naturally, there is a certain {\it gap} in the frequency spectrum near the top of the
potential hill. According to Table 1, this gap takes place between energy
levels $E_{5}$ and $E_{6}$.

\begin{table}
\caption{Dependence of the frequency $\omega_{j}$ on the energy level $E_{j}$ for
the Lennard-Jones model.}
\renewcommand{\tabcolsep}{0.12\linewidth}
\label{t1}

\begin{tabular}{|c|c|c|}
  \hline
$j$  &  $E_j$  &  $\omega_j$ \\
  \hline
1   &  -0,\,3430   &  2,\,7070 \\
  \hline
2   &  -0,\,3430   &  2,\,5711 \\
  \hline
3   &  -0,\,3410  &   2,\,4011 \\
  \hline
4  &   -0,\,3400  &   2,\,1610 \\
  \hline
5   &  -0,\,3390  &   1,\,5942 \\
  \hline
6  &   -0,\,3385   &  0,\,9243 \\
  \hline
7   &  -0,\,3374  &   1,\,1551 \\
  \hline
8   &  -0,\,3363  &   1,\,2867 \\
  \hline
9   &  -0,\,3352  &   1,\,3872 \\
  \hline
10  &   -0,\,3341  &   1,\,4710 \\
  \hline
\end{tabular}
\end{table}

\section{Stability of the {\large$\pi$}-mode vibrations near the new equilibrium positions  \label{sec_Stability}}

In previous sections, we have solved the problem of existence of the new
equilibrium positions for strained carbon chains in the framework of the DFT
ab~initio approach, as well as for the Lennard-Jones model. Now let us
consider the problem of their stability. This problem can be formulated as
follows. We must verify if a given new equilibrium position corresponds to a
minimum of the potential energy $U(\vec{X})$, which is a function of
$N$-dimensional vector~$\vec{X}=[x_{1},\, x_{2}, ... , x_{N}]$ of
{\it all degrees} of freedom of our chain. Note, that the minimum in Fig.5 corresponds to the
one-dimensional function $u(a)$, where $a$ is the $\pi$-mode amplitude, and the
vector $\vec{X}( t )=[ a,-a| a,-a
| ...  | a, -a ]$ determines the position of this
minimum in the $N$-dimensional space.

For example, below we study stability of the new equilibrium positions for
the carbon chain of $N=8$ atoms. Let us again consider the point B at the
bottom of the right potential well in Fig.~\ref{fig5}. If B represents a minimum of
the potential energy $U(\vec{X})$ in the full eight-dimensional space of
all possible displacements, then infinitesimal shift from this point in {\it any direction}
leads to increase of the function $U(\vec{X})$. It is well known that the
local extremum of the function $U(\vec{X})$ in many-dimensional space can
be of different type (minimum, maximum or a saddle point). The most
effective way to analyze the type of a given extremum point $\vec{X}_{0}$
(it determines the equilibrium position) can be carried out for the L-J
model by the following procedure.

Let us expand the function $U(\vec{X})$ near the point
$\vec{X}_{0}$ into many-dimensional Taylor series and restrict it by
quadratic terms only (this procedure corresponds to analyzing the harmonic
approximation in the framework of the dynamical approach). The obtained
{\it quadratic form} is then transformed by a certain linear transformation of variables to the {\it canonical}
form, which represents a superposition only of squares of new variables with
some coefficients $\lambda_{j}(j=1..N)$. These coefficients can be found
by diagonalization of the matrix of the original quadratic form as its
eigenvalues. The extremum $\vec{X}_{0}$ will be a minimum only if
$\lambda_{j}\ge 0$ for all $j=1..N$. If a certain eigenvalue turns out
to be negative ($\lambda_{j0}< 0$), then any infinitesimal shift
by $\gamma$ from the point $\vec{X}_{0}$ along the line $\vec{X}=\vec{X}_{0}+\gamma \vec{\xi}_{j0}$, where $\xi_{j0}$ is the eigenvector corresponding to
$\lambda_{j0}$, leads to decrease of the function $U(\vec{X})$
and, therefore, $\vec{X}_{0}$ is a saddle point (or maximum, if all
$\lambda_{j} \le 0$).

With the aid of the above procedure, we have obtained for the standart L-J
model the following results for the eight-particle chain under uniform
strain $\eta =15\,{\%}:\lambda_{1}=0$, $\lambda_{2}=+88,\!4$,
$\lambda_{3}=\lambda_{4}=+100,\!6$, $\lambda_{5}=+110,\!6$, $\lambda_{6}=\lambda_{7}=-12,\!2$, $\lambda
_{8}=-22,\!2$ (these values were rounded up to 1 figures after decimal
point). The obtained result is disappointing since the new EQP turns out to
be a saddle point (three $\lambda_{j}< 0$) and, therefore, this
equilibrium state is {\it unstable}.

\begin{figure}[h!] \centering
\includegraphics[width=1\linewidth]{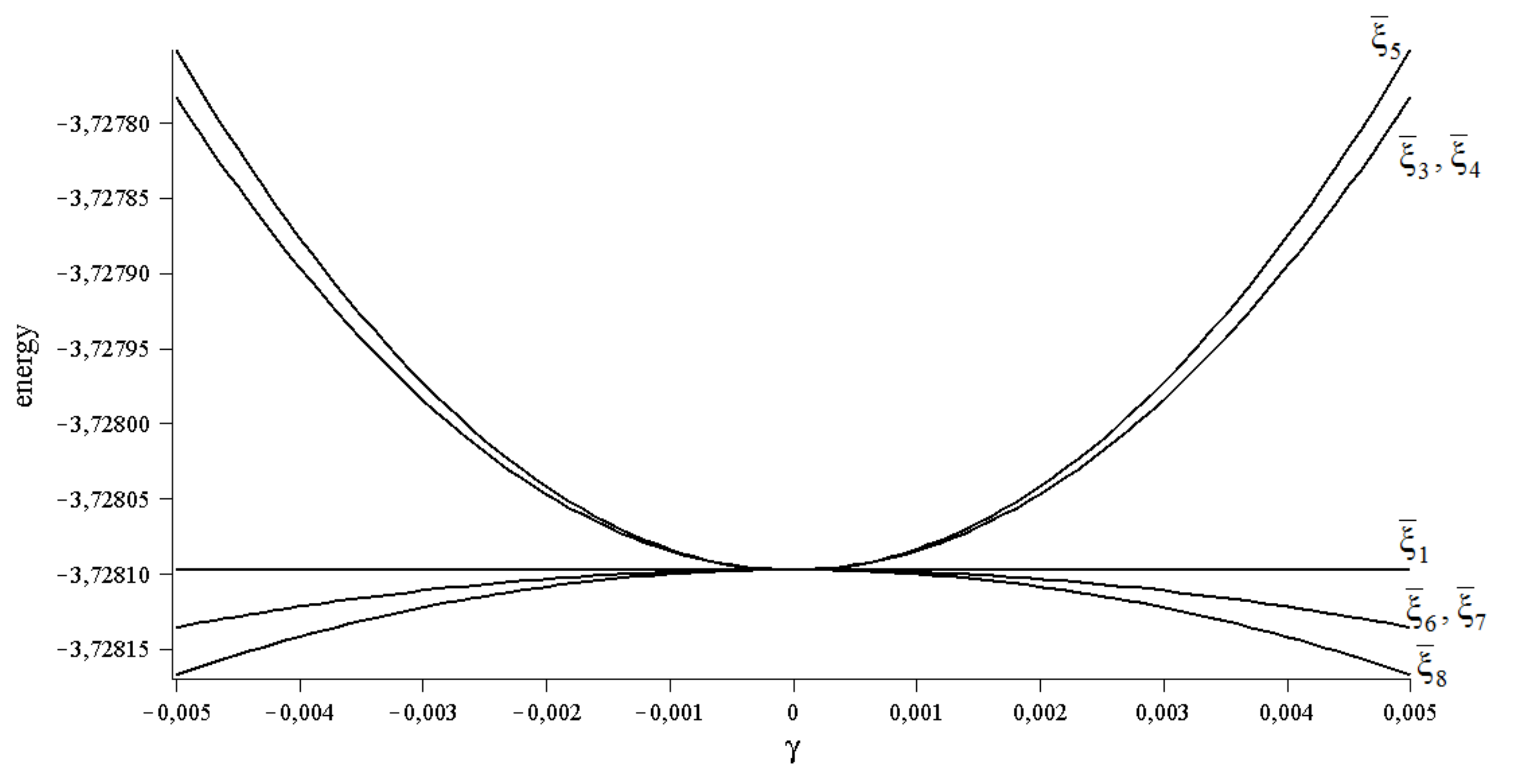} \caption{Energy profiles for the L-J chain with $N=8$ particles for all
basis directions $\vec{\xi }_{j}$~($j=1..8$) except $\vec{\xi }_{2}$. } \label{fig13}
\end{figure}

Now we can analyze the energy profiles $U(\vec{X})$ along eight lines
$\vec{X}=\vec{X}_{0}+ \gamma \vec{\xi }_{j0}$ ($j=1..8$) in the eight{-}dimensional space. Here $\gamma$ is a sufficiently small
number because we want to compare the corresponding results with those
obtained for energy function considered as a quadratic form. The energy profiles
$U(\vec{X})=u_{j}(\gamma)$ for all basis directions $\vec{\xi }_{j}$ are
depicted in Fig.~\ref{fig13}. We exclude the vector $\vec{\xi }_{2}$ from our
consideration because it determines the direction along the $\pi$-mode and,
therefore, it has no relation to the stability of this mode.

The graphs from this figure fully confirm the previously discussed results
on stability properties of the new equilibrium position $\vec{X}_{0}$.
Indeed, we see decrease of energy for $\lambda_{j}< 0$ ($\lambda
_{6}=\lambda_{7}$, $\lambda_{8})$, increase of energy for
$\lambda_{j}> 0$ ($\lambda_{3}=\lambda_{4}$,
$\lambda_{5})$ and constant energy for $\lambda_{1}=0$. Thus, the new EQP in carbyne with $15\,{\%}$ strain, which corresponds to the
vector$\vec{X}_{0}=[a_{0}, -a_{0}|a_{0}, -a_{0}|a_{0}, -a_{0}|a_{0}, -a_{0}]$ with $a_{0}=0,\!075$, occurs to be unstable in the
framework of the Lennard-Jones model. It can be proved that this conclusion
is true for arbitrary number $N$ of particles in the L-J chain.

\begin{figure}[h!] \centering
\includegraphics[width=1\linewidth]{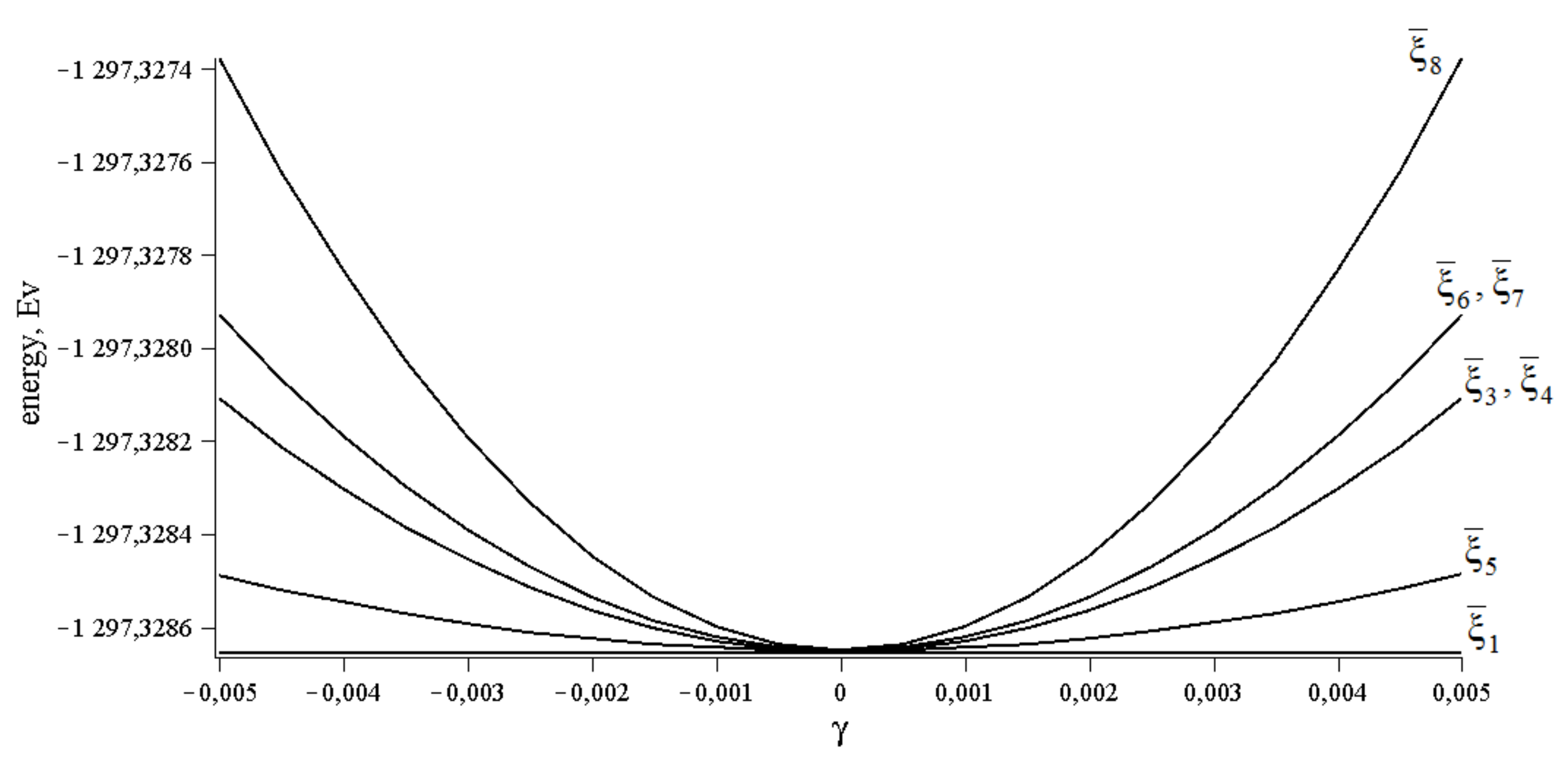} \caption{Energy profiles for the carbyne chain of $N=8$ particles for
basis directions $\mathbf{\xi }_{j}$, which are obtained in the
framework of the {\it ab initio} DFT model.} \label{fig14}
\end{figure}

The above described method can be easily applied to the study of stability
properties of new equilibrium positions in strained carbyne in the framework
of the ab initio DFT approach. In Fig.~\ref{fig14}, we present potential energy
profiles calculated by ABINIT software package \cite{PhysCommon2009,abi} for $15\,{\%}$ strained
cumulene chain of $N=8$ atoms for the above discussed basis directions.

The obtained results turn out to be unexpected, because the energy profiles
in Fig.~\ref{fig14}, unlike those in Fig.~\ref{fig13}, demonstrate {\it minima} for all basis
directions.Thus, the new equilibrium position, at point B in Fig.~\ref{fig5}, being
unstable in the L-J model, occurs to be {\it stable} in the framework of the DFT model.
Why such discrepancy does take place? We discuss this issue in the last
section of the present paper.

\begin{figure}[h!] \centering
\includegraphics[width=1\linewidth]{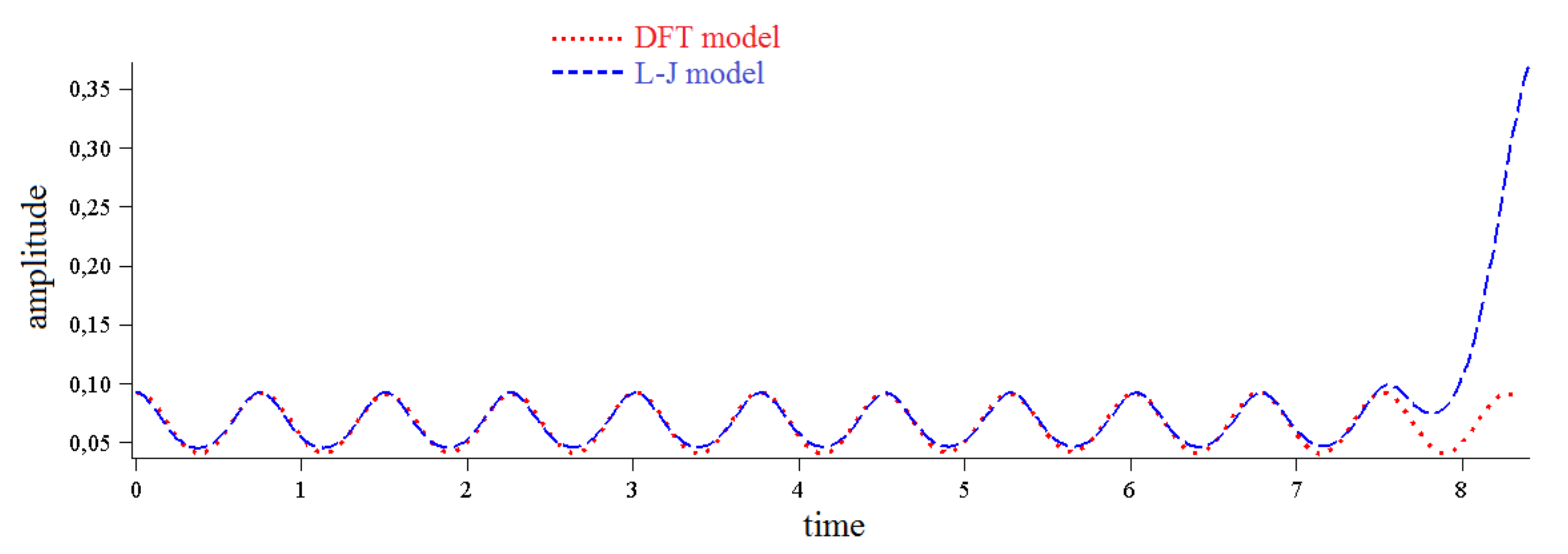} \caption{Time-evolution of the $\pi $-mode initial profile (\ref{eq5}) in DFT
and L-J models for the eight-particle carbyne chain under $15\,{\%}$ strain.} \label{fig15}
\end{figure}

Let us now consider the problem of stability of atomic {\it vibrations} near the new
equilibrium positions in strained carbon chains. In Fig.~\ref{fig15}, we represent for
$\eta =15\,{\%}$ strain the time-evolution of the carbon atom oscillations
in the right potential well (see Fig.~\ref{fig5}) approximately at the middle of its
depth. The dashed line corresponds to the DFT model, while the dotted line
demonstrates oscillations for the L-J model.

To analyze the stability of these oscillations we add to the initial $\pi
$-mode profile

\begin{equation*}
\vec{X}(0)=[ a( 0 ),-a( 0 ) | a( 0 ),-a( 0 ) |  a( 0 ),-a( 0 )  a( 0 ),-a( 0 )],
\end{equation*}
\begin{equation}
\label{eq5}
\ \ \  a(0){=0,\,075}
\end{equation}
a small perturbation in the form $\gamma \vec{\xi }_{8}$ with $\gamma
=10^{-3}$, where $\vec{\xi }_{8}$ is the basis vector, which
corresponds to the negative eigenvalue with maximal absolute value ($\lambda
_{8}={\-}22,\!2$). It is obvious from Fig.~\ref{fig15} that solution of the Newton ordinary differential
equations, corresponding to the L-J model, demonstrates {\it hard instability} (dashed line)
already at ten oscillation periods ($t\approx 10T)$. In contrast to this
result, we have found that the DFT solution for the same initial conditions
(dotted line) demonstrates a fine stability, at least up to the time
interval $t\approx 100 T$. The small modulation of the periodic oscillations
in the DFT model is induced by the above mentioned perturbation $\gamma
\vec{\xi }_{8}$ ($\gamma =10^{-3})$ artificially added to exactly
periodic initial $\pi$-mode profile (\ref{eq5}).

Thus, we can conclude that the $\pi$-mode atomic vibrations in the L-J
model are unstable, while they demonstrate stability in the DFT model (at
least, by visual inspection). Remember, that the analogical discrepancy
between static properties of the L-J model and the DFT model was discussed
in the previous section.

Because of importance of this conclusion, we verify by the rigorous Floquet
method that periodic atomic vibrations in the L-J chain are indeed unstable.
The similar stability analysis for DFT model is yet not carried out because
of some computational difficulties.

\section{Rosenberg nonlinear normal modes  \label{Ros_sec}}

Nonlinear normal mode by Rosenberg (NNM) represents a periodic vibrational
regime for which all degrees of freedom $x_{i}(t)$, at any time $t$, are
proportional to each other \cite{ApplMech1962,  Vakakis1996}. This definition can be written in the
form
\begin{equation}
\label{eq6}
x_{i}( t )= a_{i}f( t )  \ \ \ ( i = 1..N ),
\end{equation}
where $a_{i}$ are constant coefficients, while $f(t)$ is
the same time-dependent function for all degrees of freedom. If the explicit
form of dynamical equations is known, the substitution of the ansatz (\ref{eq6})
into these equations leads to a system of ($N-1$) nonlinear algebraic (or
transcendental) equations for coefficients $a_{i}$ and a single differential
equation for the function $f(t)$. Note that the
conventional linear normal modes represent a special case of Rosenberg
modes. In this case, $f(t)= \cos(\omega t+\varphi_{0})$ where $\omega $ and $\varphi_{0}$ are
frequency and initial phase, while coefficients  $\{ a_i | i=1..N\}$
 from Eq.(\ref{eq6}) are amplitudes of the
different degrees of freedom.

In contrast to the case of linear normal modes, the number of Rosenberg
modes has no relation to the dimension of the system. It is essential that
Rosenberg NNMs can exist only in very specific dynamical systems, in
particular, in those, whose potential energy is a homogeneous function of
all its arguments. On the other hand, it was shown in \cite{PhysD2005, PhysD1998} that there
can be some {\it symmetry-related} reasons for existence of NNMs in systems with arbitrary
interparticle interactions. These dynamical objects we call Rosenberg
symmetry-determined nonlinear normal modes (below the specification
``symmetry-determined'' is omitted because we consider only such type of
modes).

Let us consider $N$-particle monoatomic chain with periodic boundary conditions
and arbitrary interparticle interactions. The group theoretical method for
obtaining all nonlinear normal modes in such chain can be outlined as
follows (see details in \cite{PhysD2005}).

All dynamical regimes in the physical system with discrete symmetry group
$G_{0}$ in its equilibrium state can be classified by subgroups $G_{j}$ of
this group ($G_{j}\subset G_{0})$. One can obtain the general form of
the dynamical regime corresponding to the subgroup $G_{j}$ by solving the
equation $\hat{G}_{j} X=X$, where the vector $X(t)=
[x_{1}(t), x_{2}(t), ... , x_{N}(t)]$ describes displacements
of all degrees of freedom $x_{i}(t)$ at arbitrary fixed time $t$. In other
words, $X(t)$ must be an {\it invariant vector} of the operator group $\hat{G}_{j}$
acting in the $N$-dimensional space which is induced by the group $G_{j}$.

\begin{figure}[h!] \centering
\includegraphics[width=1\linewidth]{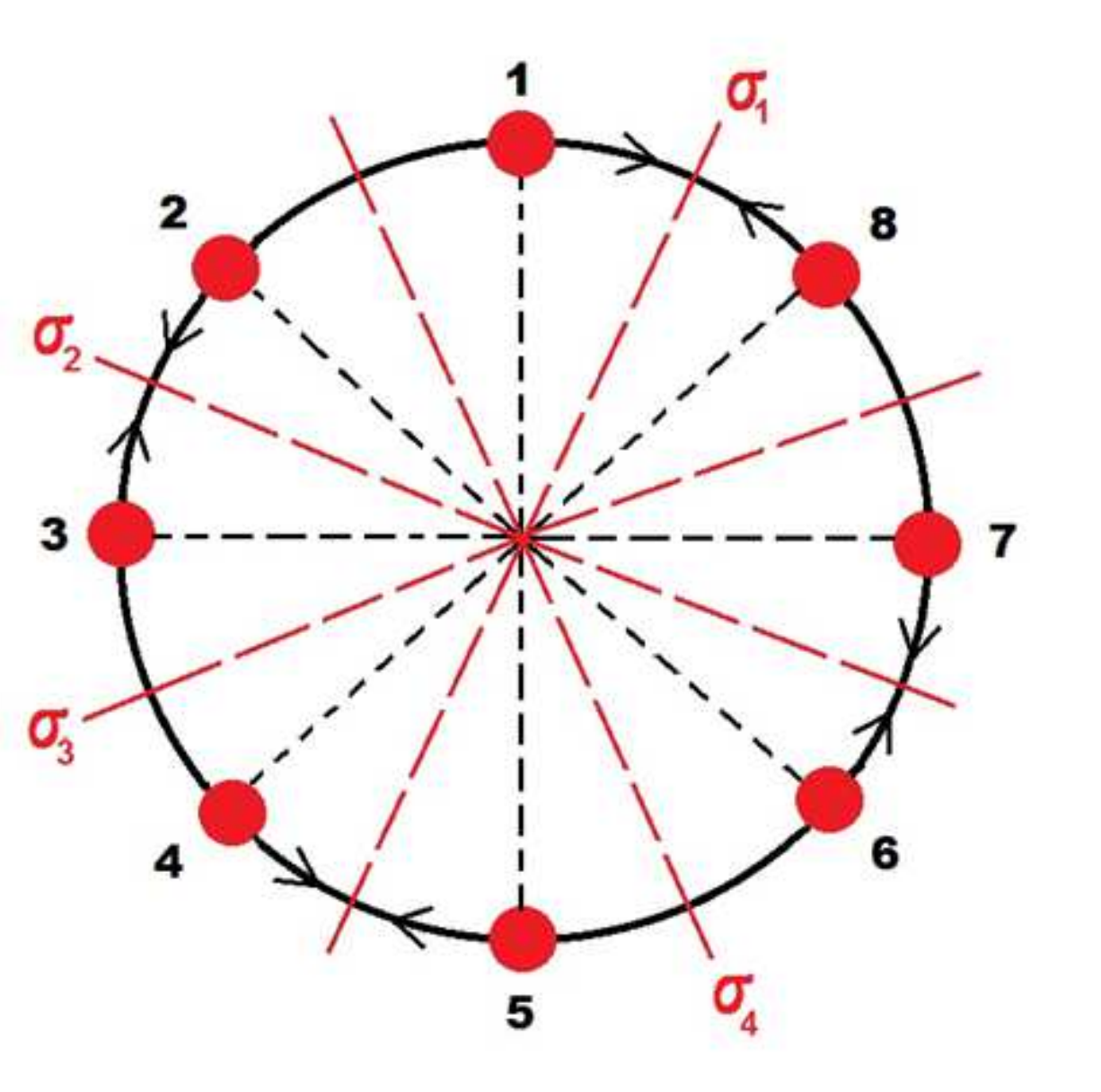} \caption{Displacement pattern of the $\pi$-mode for the chain with $N=8$ particles (color online).} \label{fig16}
\end{figure}

This chain in the equilibrium state can be represented as a ring (see, for
example, Fig.~\ref{fig16}) with the point group $D_{N}$. In \cite{PhysD2005}, we proved that only
following three Rosenberg NNMs can exist in such chains depending on the
number $N $ of their particles:
\begin{equation*}\label{eq7.1}
\pi-\mbox{mode:} \ \vec{X}(t)=[a(t), -a(t) \ | \ a(t), -a(t) \ | \ ...\ \ \ \ \ \ \ \ \ \ \ \
\end{equation*}
\begin{equation*}\label{eq7.2}
...\ | \ a(t), -a(t) \ | \ ] \ \ \mbox{for}\  \mbox{even} \ N \ (N \mbox{mod} \ 2 =0),
\end{equation*}
\begin{equation*}\label{eq7.3}
\sigma-\mbox{mode:} \ \vec{X}(t)=[a(t), 0, -a(t) \ | \ a(t), 0, -a(t) \ | \
... \ \ \ \ \ \
\end{equation*}
\begin{equation*}\label{eq7.4}
... \ | \ a(t), 0, -a(t) \ | \ ] \ \ \mbox{for } N \mbox{ mod } 3 =0,
\end{equation*}

\begin{equation*}
\tau-\mbox{mode:} \ \vec{X}(t)=[a(t), 0, -a(t), 0 \ | \ a(t), 0, -a(t),0 \ | \ ...
\end{equation*}
\begin{equation}\label{eq7}
... \ | \ a(t), 0, -a(t), 0 \ | \ ] \  \  \mbox{for} \ N \ \mbox{mod} \ 4 =0.
\end{equation}

These modes correspond to the subgroups $G_{2}=D_{N/2}$, $G_{3}=D_{N/3}$ and $G_{4} =D_{N/4}$ of the point group $G_{0}=D_{N}$,
respectively. The unit cells of the vibrational regimes (\ref{eq7}) are larger 2, 3
and 4 times, respectively, than that of the equilibrium state ($l_{0}$). In
other words, the {\it translational} symmetry of the original chain decreases by two, three or
four times if we pass from the chain equilibrium state to its vibrational
states corresponding to the $\pi$-, $\sigma$- or $\tau$-modes. Obviously, in the case of small amplitudes the above
considered Rosenberg modes tend to the linear normal modes with wave vectors
(wave numbers) $k=b/2$ ($\pi$-mode), $k=b/3$ ($\sigma$-mode), $k=b/4$ ($\tau$-mode), where $b$ is the period of the reciprocal lattice of the chain.

In dynamical simulations, above Rosenberg modes can be excited by the
following initial conditions:
\begin{equation*}
\pi-\mbox{mode:} \  \vec{X}(0)=[a, -a \ | \ a,-a \ | \ ... \ | \ a,-a], \ \ \ \ \ \
\end{equation*}
\begin{equation*}
\sigma-\mbox{mode:} \ \vec{X}(0)=[a, 0, -a \ | \ a, 0, -a \ | \ ... |  a, 0, -a],
\end{equation*}
\begin{equation*}
\tau-\mbox{mode:} \ \vec{X}(0)=[a, 0, -a, 0 \ | \ a, 0, -a, 0 \ | \ ... \ \ \ \ \ \
\end{equation*}
\begin{equation*}
... \ | \  a, 0, -a, 0],
\end{equation*}
where $a$ is an arbitrary value that determines the mode amplitudes, while the
initial velocities of all particles of the chain are equal to zero $\vec{\dot{X}}(0)=[0, 0, 0, ... , 0]$. Note that $\pi$-, $\sigma$- and $\tau$-modes represent periodic dynamical regimes determined by only one
parameter $a$.

The above presented displacement patterns (\ref{eq7}) of $\pi $-, $\sigma$- and
$\tau$-modes can be derived by demanding {\it invariance} of the general pattern $X(t)$
according to the groups $D_{N/2}$, $D_{N/3}$ and $D_{N/4}$ which are
subgroups $G_{j}$ of the group $G_{0}=D_{N}$ for the appropriate $N$. If we demand such invariance according to the group of {\it less }symmetry than the
above listed subgroups $D_{N/2}$, $D_{N/3}$ and $D_{N/4}$ (for example, for
$D_{N/5}$, $D_{N/6}$, etc.) the displacement pattern found from the
condition $\hat{G}_{j} \vec{X}=\vec{X}$ will depend not only on
{\it one} parameter $a$, but on two or more ($m> 1$) arbitrary parameters. As a
result, we obtain in such a way (see details in \cite{PhysD2005}) dynamical object that
determines {\it quasiperiodic} motion with m different basis frequencies of the Fourier
spectrum. Such patterns can be expressed as a {\it superposition} of m different NNMs and they
represent {\it bushes} of nonlinear normal modes \cite{PhysD1998, PhysDokl1993} with dimension $m>
1$. This is a reason why only three symmetry-determined Rosenberg NNMs can
exist in monoatomic chains with arbitrary interatomic interactions.

Let us note that all types of symmetry-determined periodic and quasiperiodic
vibrations in physical systems with discrete symmetries can be obtained with
the aid of specific group-theoretical methods developed in the theory of
{\it bushes} of nonlinear normal modes \cite{PhysD1998, PhysDokl1993}. The application of this theory to
various mechanical systems can be found in \cite{NonMech2003, PRE2006, PRE2004, Shcher2015, PRE2015, PRE2011, LettM2016}.

Hereafter, we write $\pi$-, $\sigma$- and $\tau$-modes in the brief form
\begin{equation*}
\pi-\mbox{mode:} \ \vec{X}=[a, \ -a], \ \ \ \ \ \ \
\end{equation*}
\begin{equation*}
\sigma-\mbox{mode:} \  \vec{X}=[a, \ 0, \ -a], \ \ \ \
\end{equation*}
\begin{equation*}
\tau-\mbox{mode:} \ \vec{X}=[a, \ 0, \ -a, \ 0],
\end{equation*}
pointing atomic displacements only in one unit cell of the vibrational
state.

\begin{figure}[h!] \centering
\includegraphics[width=1\linewidth]{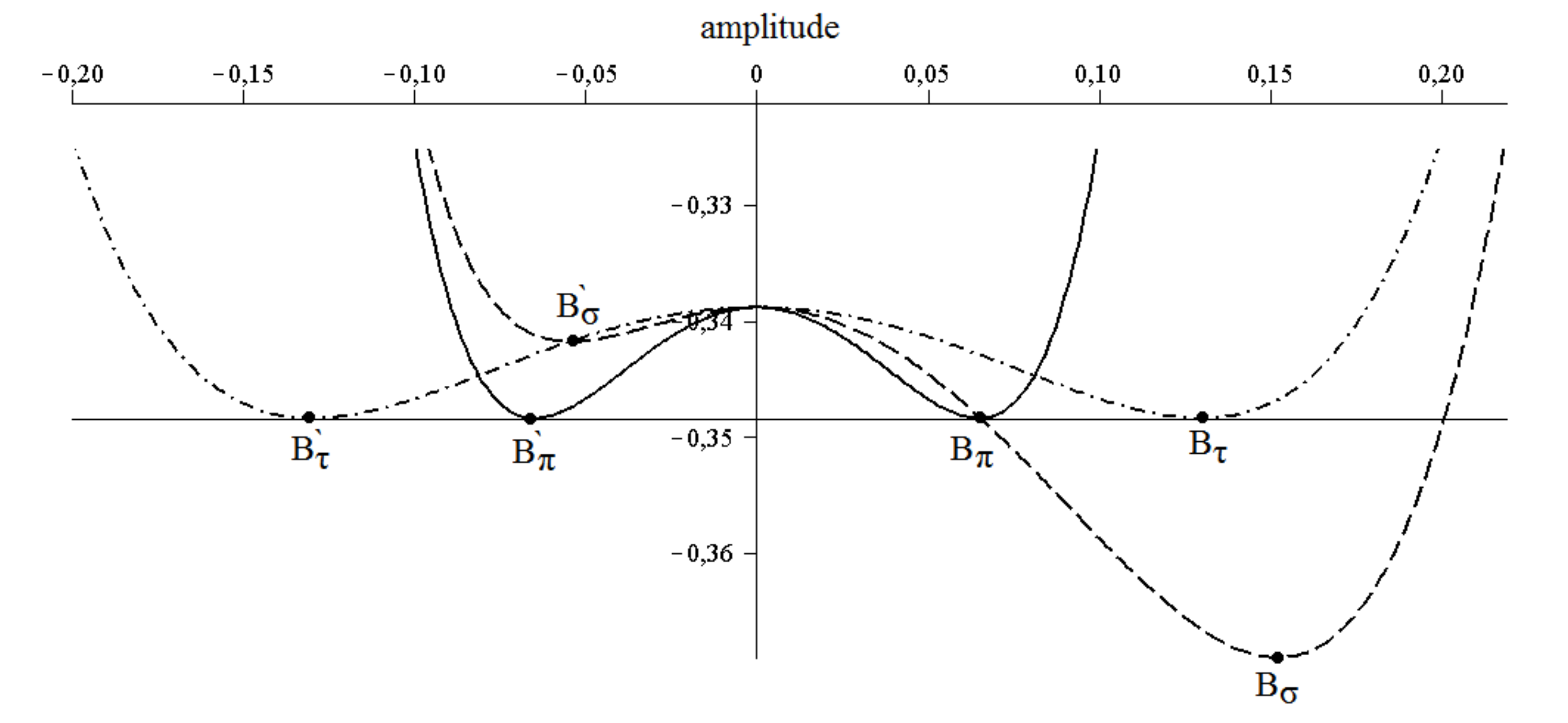} \caption{Energy profiles of $\pi$-, $\sigma$- and $\tau$-modes for
$15{\%}$ strained carbon chain with $N=12$ particles in the framework of the
Lennard-Jones model.} \label{fig17}
\end{figure}

\section{Properties of nonlinear normal modes in the framework of the Lennard-Jones model  \label{LJ_NNM_sec}}

It was shown in Sec.3 for the case of the $\pi$-mode that the L-J model
allows one to describe dynamical and static properties of strained carbyne
quite well. Let us consider applicability of this model for studying two
other Rosenberg modes in the carbon chain with $N$ atoms. Below, we present
detailed analysis for the case $N=12$ because in such chain all three
Rosenberg modes, $\pi$, $\sigma$ and $\tau$, can be exited by the
appropriate initial conditions used in the computer simulations (periodic
boundary conditions are assumed).

In Fig.~\ref{fig17}, we depict potential energy profiles for $\pi$-, $\sigma$- and
$\tau $-modes in the L-J chain strained by 15{\%}. They are calculated for
the ``standard'' Lennard-Jones potential, i.e. for the case where both
phenomenological constants of the potential (\ref{eq4}) are equal to unity
($A=B=1$). All our results are given in the corresponding dimensionless
units.

It is obvious from Fig.~\ref{fig17} that new equilibrium positions near the old one
appear for all three Rosenberg modes. The points $B_{\pi }$, $B_{\sigma }$,
$B_{\tau }$ correspond to $a> 0$, while $B'_{\pi }$,
$B'_{\sigma}$, $B'_{\tau}$ correspond to
$a< 0$. The profile of the $\sigma$-mode is asymmetrical according to
the origin $a=0$, while profiles of $\pi$- and $\tau$-modes are
symmetrical and their minima are equal to each other. As a consequence, the
equal height of potential barriers corresponds to $\pi$- and $\tau
$-modes ($\Delta u_{\pi }=\Delta u_{\tau }=0.005$). The height
of the potential barrier of the $\sigma$-mode for the right well is equal
to $0,\!015$ that approximately thrice larger than that for $\pi$- and $\tau$-modes.
The new equilibrium position ($a_{0\tau })$ for $\tau $-mode
are twice more distant from the origin ($a=0$) than that ($a_{0\pi})$ for the
$\pi $-mode (see points $B_{\tau}$ and $B_{\pi}$ in Fig.~\ref{fig17}).

We can explain the above properties of the potential profiles of the
considered Rosenberg modes in the following manner.

\begin{figure}[h!] \centering
\includegraphics[width=1\linewidth]{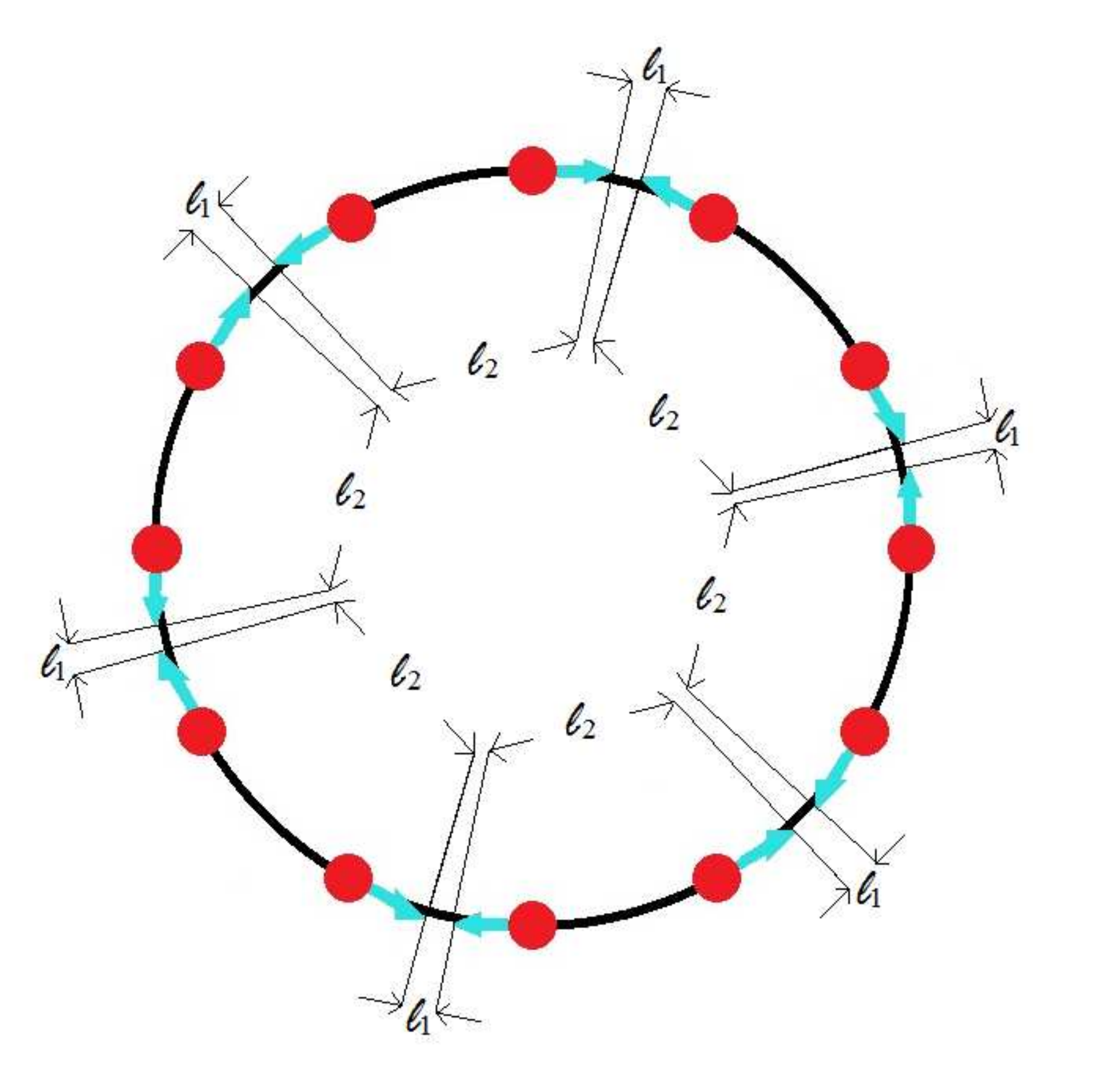} \caption{Bond lengths alternation for the $\pi$-mode structure in the Lennard-Jones chain with $N=12$ particles. Here $l_1=l_0-2a$, $l_2=l_0+2a$.} \label{fig18}
\end{figure}

\begin{figure}[h!] \centering
\includegraphics[width=1\linewidth]{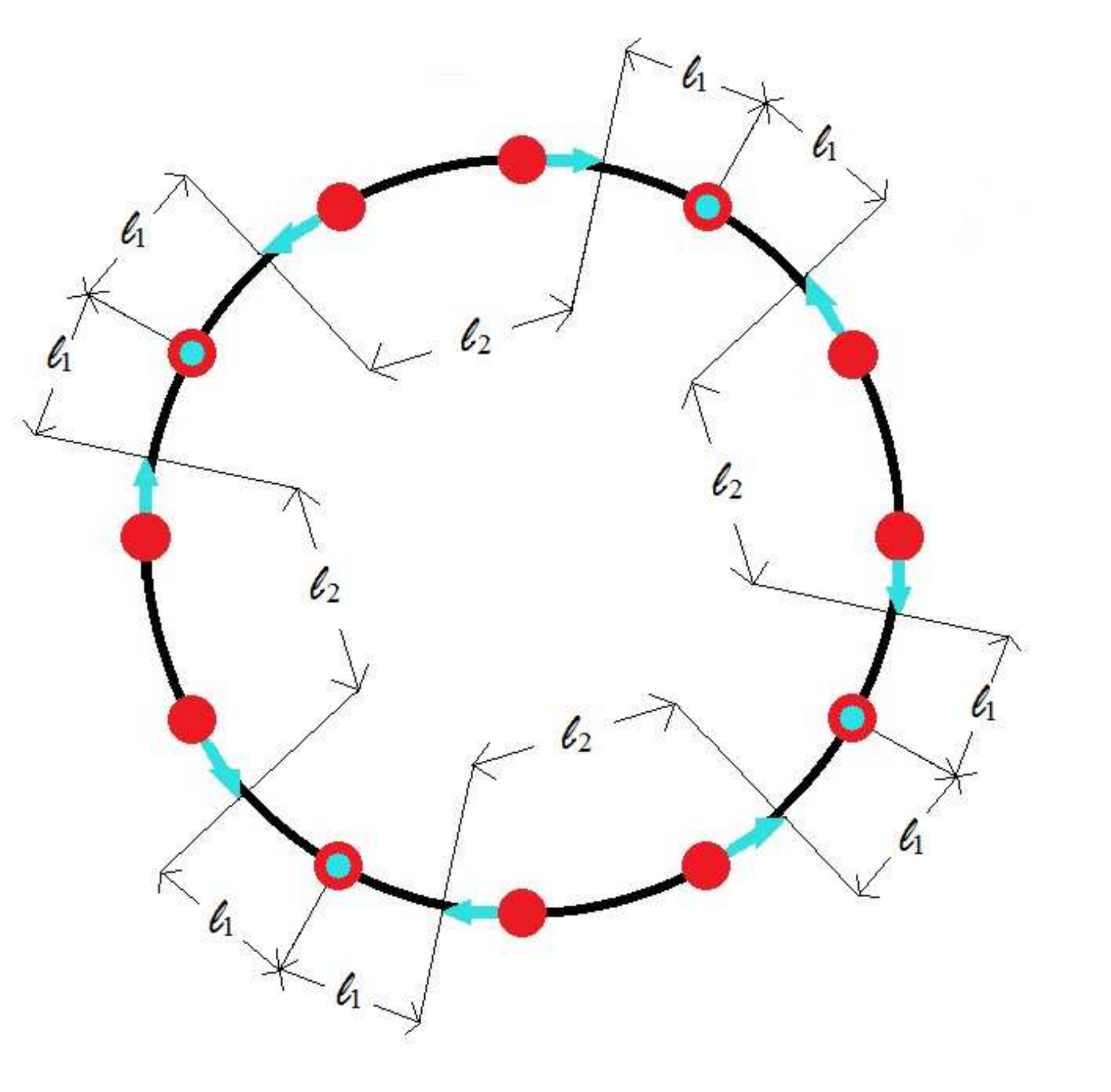} \caption{Bond lengths alternation for the $\sigma$-mode structure in the Lennard-Jones chain with $N=12$ particles. Here $l_1=l_0-a$, $l_2=l_0+2a$.} \label{fig19}
\end{figure}

\begin{figure}[h!] \centering
\includegraphics[width=1\linewidth]{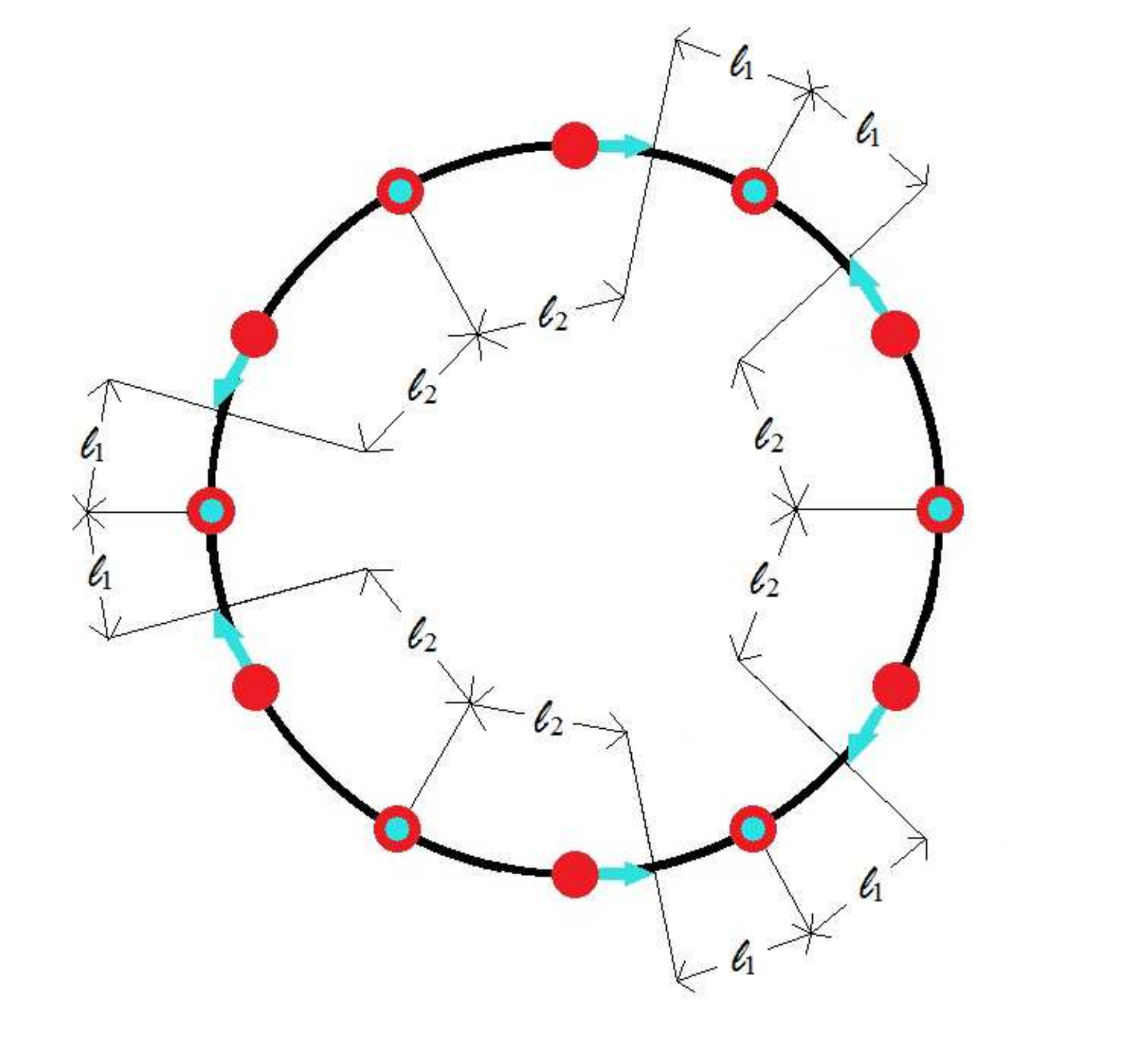} \caption{Bond lengths alternation for the $\tau$-mode structure in the Lennard-Jones chain with $N=12$ particles. Here $l_1=l_0-a$, $l_2=l_0+a$.} \label{fig20}
\end{figure}

Let us consider Figs.17-19. We depict there L-J chains as rings of
$N=12$ particles for a fixed parameter $a$ (amplitude) of $\pi$-, $\sigma$-
and $\tau$-modes, respectively. The atomic displacements are shown by
arrows. In each of these figures, one can see {\it two types} of interatomic distances
(bond lengths), $l_{1}$ and $l_{2}$. For the $\pi$-mode,
$l_{1}=l_{0}-2a$ and $l_{2}=l_{0}+2a$, where $l_{0}$ is the
interatomic distance of equilibrium state of the strained chain. For
$a> 0$, $l_{1}$ represents the short bond, while $l_{2}$
represents long one, and vice versa for $a< 0$. Thus the difference
$\Delta l=| l_{1}-l_{2} |$ between the interatomic
distances turns out to be $\Delta l=4 | a | $. According to
Fig.~\ref{fig18} the following sequence of the bond lengths (BLs) takes place
for the $\pi$-mode:

\begin{equation}\label{eq8}
BLA_{\pi }=[l_{1}l_{2} \ | \ l_{1}l_{2} \ | \ l_{1}l_{2} \ | \ l_{1}l_{2} \ | \ l_{1}l_{2} \ | \
l_{1}l_{2}].
\end{equation}

Let us consider a pair potential $\varphi (r)$, which describes interactions
between neighboring particles placed at distance r from each other. There
are six bond lengths (BLs) of $l_{1}$ type and six BLs of $l_{2}$ type in
Eq.~(\ref{eq8}). Therefore, the whole potential energy $U(a)$ of the considered chain of
$N=12$ particles is equal to

\begin{equation}\label{eq9}
U_{\pi }(a)=6\varphi (l_{1})+6\varphi (l_{2}),
\end{equation}
where $l_{1}=l_{0}-2a$, $l_{2}=l_{0}+2a$. The position of the energy
minima is determined by the equation

\begin{equation}\label{eq10}
 dU_{\pi }(a)/da=0
\end{equation}
from which we find

\begin{equation}\label{eq11}
\varphi '(l_{0}+2a)=\varphi'(l_{0}-2a).
\end{equation}
If $x_{0}> 0$ represents the root of the equation \footnote{In
Sec.~\ref{LJ_sec}, we have solved this equation for the Lennard-Jones potential by
simple graphic method.}

\begin{equation}\label{eq12}
\varphi'(l_{0}+x)=\varphi'(l_{0}-x),
\end{equation}
then the new equilibrium position, corresponding to the bottom of the right
potential well (point $B_{\pi }$ in Fig.~\ref{fig17}) for the L-J chain with 15{\%}
strain, is equal to $a_{0}=0,\!075$ and, therefore, $\Delta
l=4a_{0}=0,\!30$.

Using the same method for the $\tau$-mode (Fig.~\ref{fig20}) in the L{-}J~chain, we
obtain

\begin{equation}\label{eq13}
l_{1}=l_{0}-a, \ \ l_{2}=l_{0}+a, \ \ \Delta l=2a.
\end{equation}

\begin{equation}\label{eq14}
BLA_{\tau }=[l_{1}l_{1}l_{2} l_{2} |
l_{1}l_{1}l_{2}l_{2} | l_{1}l_{1}l_{2} l_{2}],
\end{equation}

\begin{equation}\label{eq15}
	U_{\tau}(a)= 6\varphi (l_{1})+6\varphi (l_{2}),dU_{\tau}(a)/da=0,
\end{equation}

\begin{equation}\label{eq16}
\varphi'(l_{0}+a)=\varphi'(l_{0}-a).
\end{equation}

Therefore, $a_{0\tau }=x_{0}$ where $x_{0}$ is the root of Eq.~\ref{eq12}.
Comparing this result with that for the $\pi$-mode, $a_{0\pi }=x_{0}/2$,
we conclude that $a_{0\tau } = 2a_{0\pi }$, i.e. potential minimum for $\tau$-mode
for {\it every} pair interparticle potential $\varphi (x)$ (not only for the
Lennard-Jones potential!) is twice more distant from the origin at $a=0$.

Let us now consider Fig.~\ref{fig19} corresponding to the $\sigma$-mode. We find from
this figure for $a> 0$.

\begin{equation}\label{eq17}
l_{1}=l_{0}-a, \ \ l_{2}=l_{0}+2a, \ \ \Delta l=3a,
\end{equation}

\begin{equation}\label{eq18}
BLA_{\sigma }=[l_{1}l_{1}l_{2} |
l_{1}l_{1}l_{2} | l_{1}l_{1}l_{2} |
l_{1}l_{1}l_{2}],
\end{equation}

\begin{equation}\label{eq19}
U_{\sigma }(a)=8\varphi (l_{1})+4\varphi (l_{2}).
\end{equation}

Thus, equation $dU_{\sigma }(a)/da=0$ for finding equilibrium position
$a_{0\sigma}$ looks as follows

\begin{equation}\label{eq20}
\varphi'(l_{0}+2a)=\varphi' (l_{0}-a).
\end{equation}

Note that this equation differs essentially from the Eqs.~(\ref{eq11}) and (\ref{eq16}) for
$\pi$- and $\tau$-modes. Indeed, both equations (\ref{eq11}) and (\ref{eq16}) are
invariant under changing the sign of the parameter $a$. In contrast to that,
Eq.~(\ref{eq20}) changes its form under the above transformation $(a)\rightarrow(-a)$ as
follows:

\begin{equation}\label{eq21}
\varphi'(l_{0}-2a)=  \varphi'(l_{0}+a).
\end{equation}

Let us note that Eqs. (\ref{eq16}) and (\ref{eq20}) possess {\it different} roots. As a consequence, we
obtain for the L-J model under 15{\%} strain (see Fig.~\ref{fig17}) \ $B_{\sigma }=0,\!150$,
while $B'{\sigma}=-0,\!075$.

The matter is that transformation $(a)\rightarrow (-a)$ changes the number of short
and long interatomic distances (bond lengths), because $l_{1}=l_{0}-a$ is
a short distance and $l_{2}=l_{0}+2a$ is a long distance for
$a> 0$, with the opposite situation for $a< 0$. For $N=12$
particles chain, there are 8 short bonds and 4 long bonds for the case
$a> 0$, while 4 short and 8 long bonds correspond to the case
$a< 0$. This is the cause of the asymmetrical form of the potential
profile of the $\sigma$-mode. (For the case of $\pi$- and $\tau$-modes
the transformation $(a)\rightarrow(-a)$ remains the number of short and long bonds
unchanged, they simply are transformed into each other
$l_{1}\leftrightarrow l_{2})$.

Thus, condensation of $\pi$-, $\sigma $- and $\tau $-modes results in
different schemes of bond lengths alternation:

\begin{equation}\label{eq22}
BLA_{\pi }=[l_{1}l_{2} | l_{1}l_{2} |
l_{1}l_{2} | l_{1}l_{2} | l_{1}l_{2} |
l_{1}l_{2}],
\end{equation}

\begin{equation}\label{eq23}
 BLA_{\sigma }=[l_{1}l_{1}l_{2} |
l_{1}l_{1}l_{2} | l_{1}l_{1}l_{2} |
l_{1}l_{1}l_{2}],
\end{equation}

\begin{equation}\label{eq24}
 BLA_{\tau }=[l_{1}l_{1}l_{2}l_{2} |
l_{1}l_{1}l_{2}l_{2} | l_{1}l_{1}l_{2}l_{2}].
\end{equation}

It seems to be plausible that there can exist two {\it new forms} of carbyne besides
cumulene and polyyne. Indeed, all bonds are equal in cumulene
[$l_{1}l_{1}l_{1}... l_{1}$], while there are two different bonds in
polyyne and in carbynes appearing as a result of condensation of $\sigma$-
or $\tau$-modes, but they possess {\it different alternation schemes} as given by Eqs. (\ref{eq22}) - (\ref{eq24}).

The present experimental technique cannot allow one to find certainly the
alternation of bond lengths in carbon chains. However, let us note that in
\cite{Carb2014} the rough attempt of finding bond lengths in strained carbon chain was
realized by transmission electron microscope (TEM). One can hope that
progress in the experimental technique will soon allow determining bond
lengths alternation for strained carbyne with reasonable accuracy.

In conclusion, it is interesting to note that BLA depicted in Fig.1 of the
paper \cite{Nano2013}, which was found by DFT simulations, is more close to the
condensation of Rosenberg modes different from the $\pi$-mode, which
corresponds to polyyne form of carbyne.

Now we consider the critical value $\eta_{c}$ of the strain for which old
equilibrium positions of the carbon atoms in the chain with arbitrary pair
potential $\varphi (r)$ lose their stability and new such positions appear.
The potential energy $u(a)$ of the chain in the small vicinity of the parameter
$a=0$ has to possess minimum for $\eta <\eta_{c}$ and maximum
for $\eta >\eta_{c}$.

Let us write, the first terms of the Taylor series of the function $u(a)$ near
the point $a=0$ for the $\pi$-mode [see Eq.(\ref{eq9})]:

\begin{equation*}
 u_{\pi }(a)= 6\varphi (l_{0}-2a)+ 6\varphi (l_{0}-2a)=
12\varphi (l_{0})+
\end{equation*}
\begin{equation}\label{eq25}
+ 24a^{2}   \varphi''(l_{0})+...
\end{equation}

Obviously, maximum of the function $u_{\pi }(a)$ takes place if $\varphi
''(l_{0})< 0$ and minimum if $\varphi
''(l_{0})> 0$. Therefore, the critical value of the strain
$\eta_{c}$ can be determined from the equation $\varphi
''(l_{0})=0$, where $l_{0}= l_{0}(\eta)$ is the interatomic
distance of the chain in the equilibrium state under the strain $\eta$.

Thus, $\eta_{c}$ corresponds to the point at which curvature of the
function $\varphi (r)$ changes sign. For the case of the Lennard-Jones
potential $\varphi (r)=r^{-12}- r^{-6}$, we find that $\varphi
''(r)=0$ for $r=\sqrt[6]{\frac{26}{7}}\approx 1,\!244$, while
equilibrium interatomic distance for the L-J chain without strain is equal
to $r_{0}=\sqrt[6]{2}\approx 1,\!122$. Therefore, $\eta
_{c}=(r-r_{0})/r_{0}~100{\%}\approx 10,\!87{\%}$.

Similarly to the above calculations, we obtain the following decompositions
for the $\sigma$-mode and $\tau$-mode:

$u_{\sigma }(a)=12\varphi (l_{0})+ 12a^{2}  \varphi
''(l_{0})+ ...$,

$u_{\tau }(a)=12\varphi (l_{0})+ 6a^{2} \varphi''(l_{0})+... $

Therefore, we find {\it identical} conditions for obtaining the critical value $\eta
_{c}$ for all three Rosenberg modes: $\varphi''(l_{0})=0$.

\begin{figure}[h!] \centering
\includegraphics[width=1\linewidth]{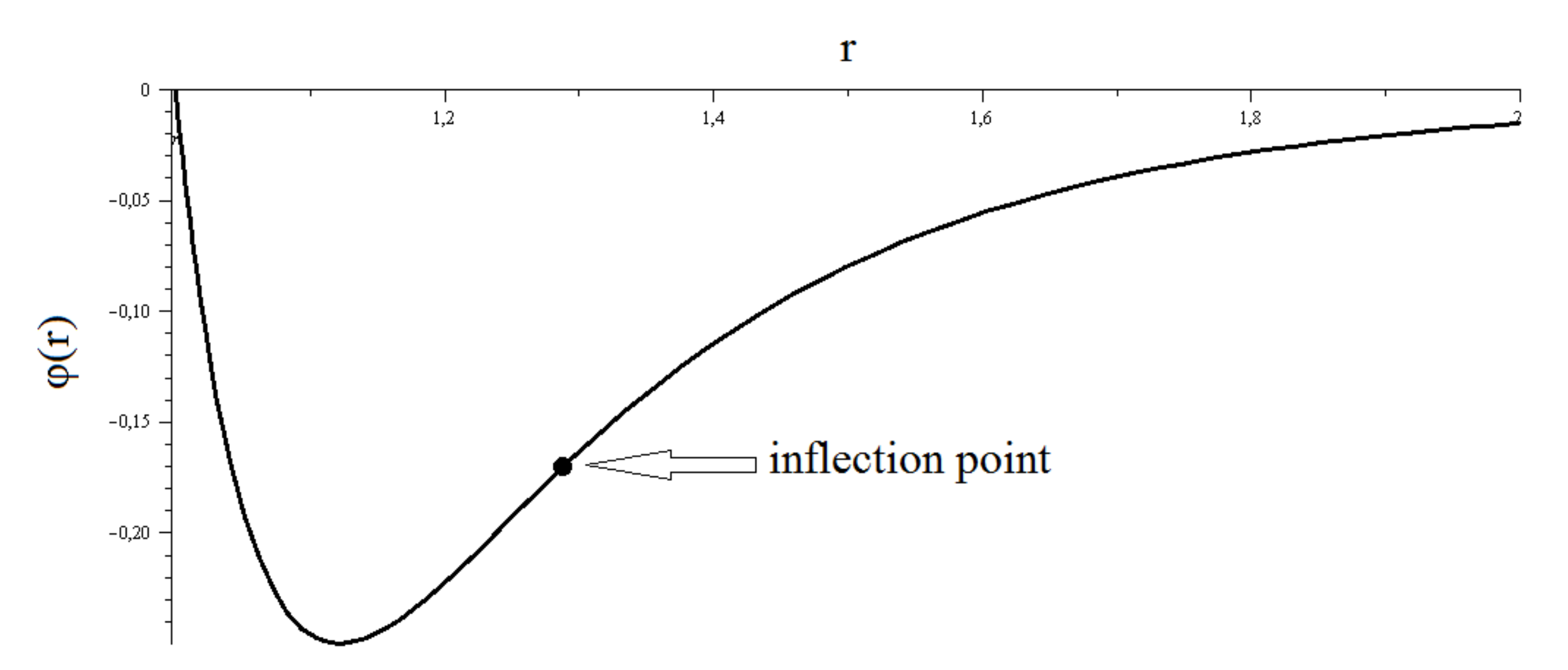} \caption{The inflection point of the standard Lennard-Jones potential.} \label{fig21}
\end{figure}

In other words, the stability bifurcation of the L-J chain with increasing
of the strain $\eta $ takes place simultaneously for these modes, i.e. for
the same $l_{0}$ corresponding to the point of zero curvature (inflection
point) of the pair potential $\varphi (r)$. This point for the Lennard-Jones
potential is shown in Fig.~\ref{fig21}.

\vspace{-10pt}

\section{Conclusion  \label{Conclus_sec}}

In this paper, we study properties of longitudinal nonlinear vibrations in
strained carbon chains by ab initio methods, based on the density functional
theory, as well as by the methods of molecular dynamics in the framework of
the simple Lennard-Jones model, and also for arbitrary pair potential of
the interparticle interactions.

In the brief paper \cite{Lett2016}, we have been reported that under sufficiently large
uniform strain ($\eta >11{\%}$) radical transformation of carbyne cumulene
structure takes place as the result of condensation of the $\pi$-mode,
which becomes a soft mode in a certain range of its amplitude. Indeed, in
this case, the old equilibrium positions, around which carbon atoms vibrate
in the case of small strain, lose their stability and two new such positions
arise near each of them. This transformation corresponds to the Peierls
phase transition in one{-}dimensional crystal and is associated with
doubling of the unit cell. As a result, the carbyne chain transforms from
the cumulene to polyyne form. This phase transition in strained carbyne
chains was previously discovered in the work \cite{NanoLett2014}, which was also carried out
by ab initio DFT simulations, but from another position compared to those
of \cite{Lett2016}. Indeed, we have studied nonlinear vibrations in carbyne chains,
while the authors of \cite{NanoLett2014} explored their static structure.

As was already emphasized in Introduction, the above phase transition is a
phenomenon of great physical importance because it leads to a qualitative
change in the electron spectrum of cumulene. Indeed, as a result of
appearance of the energy gap in this spectrum, conductive cumulene
transforms into polyyne which is semiconductor or insulator. In turn, it is
expected that such change in electrical properties under mechanical strain
can be used in some nanodevices.

In \cite{NanoLett2014}, a quantum-mechanical model of one particle moving in the double-well
potential was introduced to explain the properties of the Peierls phase
transition found in the framework of the DFT approach. Unlike this in \cite{Lett2016},
for the same purpose, a more simple classical model was introduced, which
represents the chain of mass-points whose interactions are described by the
Lennard-Jones potential (L-J chain). Using this model, we were able to
explain not only the static properties, such as appearance of the new
equilibrium positions, etc., but also the properties of nonlinear vibrations
near these new equilibrium positions. Moreover, the approach developed in
\cite{Lett2016} allows us to consider the condensation of two other symmetry-determined
Rosenberg nonlinear normal modes (besides the $\pi$-mode), which correspond
to multiplication of vibrational unit cell three- or four times compared to
that of cumulene equilibrium state. In turn, such approach leads to
prediction of possibility for existence of two new forms of carbyne with the
bond length alternations different from that of polyyne.

In dynamical regimes, described by Rosenberg modes, all particles of the
system vibrate with the same frequency, but unlike the conventional linear
normal modes, these vibrations are {\it nonlinear}. In \cite{PhysD2005}, we proved with the aid of
group-theoretical methods that in the monoatomic chains with arbitrary
interactions only three Rosenberg modes, $\pi $, $\sigma $ and $\tau $, can
exist. The unit cell of the vibrational states corresponding to $\pi$-,
$\sigma$- and $\tau$-modes are two, three and four times larger than that
of the chain equilibrium state.

The condensation of these modes induces the carbon structures with two
different bond lengths and alternation schemes of these bonds are different
for the structures induced by $\pi$-, $\sigma$- and $\tau$-modes [see
Eqs. (\ref{eq22})-(\ref{eq24})]. The condensation of the $\pi$-mode generates the
polyyne structure of the carbon chain, while condensation of $\sigma$- and
$\tau$-modes may be associated with two new forms of carbon chains.

We use two different models to simulate dynamical and static properties of
the above Rosenberg modes in strained carbon chains, the DFT model and the
Lennard-Jones model. The former model is certainly more adequate because it
takes into account the presence of electron shells near each carbon nucleus
and provides quantum-mechanical approach for the electron subsystem. However,
density functional theory cannot ensure exact consideration of the physical
reality because of a number of approximations used in practical application
of this theory. In this sense, the crucial role plays the choice of
exchange-correlation functionals. It is known that such functional
conventionally used in the software packages for DFT simulations (ABINIT,
VASP etc.) cannot ensure acceptable approximation for calculation of van der
Waals interactions \footnote{Note that the appropriate corrections to the
exchange-correlation functionals are essentially {\it nonlocal}. As a consequence the LDA
approximation and even GGA approximation cannot provide any satisfactory
calculation of vdW interactions.} (see discussion of this problem in
\cite{VDW1,VDW2,VDW3,VDW4}. On the other hand, the van der Waals (vdW) interactions turn out
to be very essential for large strain of carbyne since the distances of its
atoms are considerably greater than those in the chain without strain.

In the Lennard-Jones model, vdW interactions are taken into account by the
last term of the potential (\ref{eq2}), which is proportional to $r^{-6}$. Note that
the main problem of the choice of vdW amendments to the exchange-correlation
functionals is how to obtain their correct asymptotic ($r^{-6})$ \cite{VDW1,VDW2,VDW3,VDW4}.

We could not get an acceptable approximation in the framework of the DFT
approach to simulate correctly the condensation of $\sigma $- and $\tau$-modes (obviously, the vdW interactions turn out to be more essential for
these modes than for the $\pi $-mode\footnote{For example, the bottom of the
potential well for $\sigma $-mode and $\tau $-mode are twice more distant
from origin (see Fig.~\ref{fig16}) than that of the $\pi $-mode.}). As a consequence,
our prediction of the possibility of existence of two new forms of carbon
chains is based only on studying of the Lennard-Jones model.

 Now let us return to consideration of stability properties of the new EQPs
and of the periodic oscillations in their vicinity, which are discussed in
Secs.~\ref{LJ_sec} and \ref{sec_Stability} of this paper. We present there computational experiments
demonstrating that our Lennard-Jones model, which is rather accurate for
prediction of new static and dynamical properties of the strained carbon
chains, turns out to be unsatisfactory for analyzing stability of the new
atomic equilibrium positions and stability of oscillations around them. Now
we can indicate the following causes of such phenomenon.

As was already emphasized, the DFT model, unlike the L-J model, is more
adequate in physical sense because it takes into account electron shells of
each carbon atom, which adapt to any change in the nuclear configuration,
while the L-J model deals with bare mass points. In other words, many
degrees of freedom correspond to each site in the DFT model, while only one
variable is associated with every site in the Lennard-Jones model. There is
also a mathematical cause of the above discrepancy between DFT and L-J models. Indeed, in the case of
the L-J model we study the stability of a solution of ordinary differential
equations (classical Newton equations), while in the case of the DFT model
one has to analyze the stability of the solution to a system of cumbersome
integro-differential equations (Kohn-Sham quantum-mechanical equations).

Thus, the L-J model describes sufficiently well the {\it existence} of new equilibrium
positions in strained carbyne, but it turns out to be unsatisfactory for
analysis of their {\it stability}. We feel that such discrepancy between the results
obtained by methods of molecular dynamics and those by DFT simulations
may be typical for very different physical problems.

\begin{acknowledgments}
The authors are sincerely grateful to Profs. V.~P.~Sakhnenko,  S.~V.~Dmitriev and N.~V.~Ter-Oganessian for useful discussions, and to I.~P.~Lobzenko for assistance in application of the software package ABINIT. This work was supported by the Russian Science Foundation.
\end{acknowledgments}

\end{document}